\documentclass[twocolumn,letterpaper]{aa}
\usepackage{graphicx,psfig}
\usepackage{txfonts}
\newfont{\myfont}{cmmib10}

\DeclareSymbolFont{cmmi}{OML}{cmm}{m}{it}
\DeclareMathSymbol{v}{\mathalpha}{cmmi}{"76}
\newcommand{\viss}{v_{\rm ISS} }

\begin{document}
\title{Diffractive Interstellar Scintillation of the Quasar J1819$+$3845 at $\lambda\,21$\,cm}

   \author{J.-P. Macquart
          \inst{1,2}\thanks{Jansky Fellow}
          \and
          A.G. de Bruyn\inst{3,1}
          }

   \offprints{J.-P. Macquart}

\institute{Kapteyn Astronomical Institute, University of Groningen, P.O. Box 800, Groningen 9700 AV, The Netherlands
\and
National Radio Astronomy Observatory, P.O. Box 0, Socorro NM 87801, U.S.A.
 \email{jmacquar@nrao.edu}
 	\and 
Netherlands Foundation for Research in Astronomy, Dwingeloo, The Netherlands  \email{ger@astron.nl}
             }

   \date{Received 25 April 2005; accepted 12 September 2005}

\titlerunning{Diffractive scintillation in J1819$+$3845}
\authorrunning{Macquart \& de Bruyn}

\abstract{ We report the discovery of fast, frequency-dependent
intensity variations from the scintillating intra-day variable quasar
J1819$+$3845 at $\lambda\,21$\,cm which resemble diffractive
interstellar scintillations observed in pulsars. The observations were
taken with the Westerbork Synthesis Radio Telescope on a dozen
occasions in the period between Aug 2002 and Jan 2005.  
The data were sampled at both high temporal and 
high frequency resolution and have an
overall simultaneous frequency span of up to 600 MHz.  In constructing
the light curves and dynamic spectra the confusion from background 
sources has been eliminated.   The timescale (down to 20 min) and
the bandwidth (frequency decorrelation bandwidth of 160 MHz) 
of the observed variations jointly imply that the component of the
source exhibiting this scintillation must possess a brightness
temperature well in excess of the inverse Compton limit.  A specific
model in which both the source and scintillation pattern are isotropic
implies a brightness temperature $0.5 \times 10^{13} z_{\rm pc}\,$K,
where previous estimates place the distance to the scattering medium
in the range $z_{\rm pc}=4-12$\,pc, yielding a minimum brightness
temperature $>20$ times the inverse Compton limit.  An independent
estimate of the screen distance using the 21\,cm scintillation
properties alone indicates a minimum screen distance of $z \approx
40\,$pc and a brightness temperature above $2 \times 10^{14}\,$K.
There is no evidence for anisotropy in the scattering medium or source
from the scintillation characteristics, but these estimates may be
reduced by a factor comparable to the axial ratio if the source is
indeed elongated.  The observed scintillation 
properties of J1819$+$3845 at 21\,cm  are compared with those at 6\,cm,
where a significantly larger source size has been deduced for the bulk
of the emission by Dennett-Thorpe \& de Bruyn (2003). 
However, opacity effects within the source and 
the different angular scales probed in the regimes of weak and strong 
scattering complicate this comparison.

\keywords{Quasars: individual: J1819$+$3845 --
Galaxies: active -- Scattering -- Radiation mechanisms: non-thermal --
Techniques: high angular resolution } }

   \maketitle
%

\section{Introduction}

The presence of intra-day variability (IDV) in some active galactic
nuclei (AGN) at centimetre wavelengths (\cite{Heeschen};
\cite{Bonngp1}) raises concern that the brightness temperatures of
radio sources may violate the inverse Compton limit by several orders
of magnitude (\cite{Bonngp2}; \cite{Kedziora-Chudczer}).

It is now recognized that the variability is largely, if not
exclusively, due to scintillation in the interstellar medium of our
Galaxy (e.g. \cite{Jauncey00}; \cite{Lovell}).  This is established
unequivocally for the IDV quasar J1819$+$3845.  A $\sim\!\!90\,$s time
delay in the arrival times of the source's intensity variations
measured between two widely-separated telescopes firmly identifies its
variability with interstellar scintillation.  The finite delay is
attributed to the finite speed with which the scintillation pattern
moves transverse to the line of sight (\cite{DennettThorpe02}).
An annual modulation in the timescale of the variability is also
observed in J1819$+$3845, and is explained by the annual modulation in
the scintillation velocity, ${\bf v}_{\rm ISS}$, due to the Earth's
changing orbital velocity relative to the interstellar scattering
material (\cite{DennettThorpe03}).  This annual cycle arises because
the Earth's velocity is comparable to the velocity of the scattering
material in the interstellar medium responsible for the intensity
fluctuations.  Annual cycles are also reported in several other IDV
sources (\cite{Bignall}; \cite{Rickett01}; \cite{JaunceyM}).

The brightness temperatures of IDV sources have proven difficult to
constrain, largely because of the difficulty of determining the
distance to the scattering material, $z$, responsible for the
intensity fluctuations.  These brightness temperatures have hitherto
been estimated on the basis of the measurements made at $\lambda
6\,$cm, at which flux variations are caused by weak interstellar
scintillation.  In this regime one measures the source variability
timescale relative to the Fresnel timescale, $t_{\rm F}=r_{\rm
F}/\viss = (\lambda z/2 \pi)^{1/2}/ \viss$ and thus deduces the source
size relative to the angular scale subtended by the Fresnel scale,
$\theta_{\rm F}=r_{\rm F}/z$.  Despite these difficulties, several
sources still exceed the inverse Compton limit by a large margin.  A
thorough analysis estimates the redshift-corrected brightness
temperature of PKS\,0405$-$385 at $5 \times 10^{13}\,$K
(\cite{Rickett02}).  A similar but less robust limit of $T_b \geq 5
\times 10^{13}$\,K is also derived for the source PKS\,1519$-$273
(\cite{Macquart00}).

J1819$+$385 is estimated to possess a modest brightness temperature of
approximately $\sim\!\! 10^{12}\,$K at 4.9\,GHz (Dennett-Thorpe \& de Bruyn 2003).  The flux density of
the source varies on timescales as short as 40\,minutes at this
frequency.  These variations are also interpreted in terms of
scintillation in the regime of weak scattering
(\cite{DennettThorpe00}).  The estimated screen distance is small, in
the range $z=4-12\,$pc, and the source is relatively weak, which
accounts for the small brightness temperature despite its extremely
rapid variability (\cite{DennettThorpe03}).

This source generally exhibits slower variations at lower frequencies
(\cite{DennettThorpe00}). The dominant variations at 2.4 and 1.4\,GHz 
occur on $\sim\!\!\!6$-hour timescales during the same interval in the annual
cycle of J1819$+$3845 in which dramatic intra-hour variations are observed at
4.9\,GHz.  Such a change in the character of the variability with
decreasing frequency is typical of intra-day variable sources
(\cite{Kedziora-Chudczer}; \cite{Macquart00}; \cite{Quirrenbach00}).
These slow variations are attributed to the increase of scattering
strength with frequency, and are associated with refractive
scintillation in the regime of strong scattering.  Such scattering
occurs when the Fresnel scale $r_{\rm F}$ exceeds the diffractive
scale length, $r_{\rm diff}$, the transverse length scale over which
the mean square difference in phase delay imposed by plasma
inhomogeneities in the ISM is one radian.  Refractive variations occur
on a timescale $\sim \!  r_{\rm F}^2/r_{\rm diff} \equiv r_{\rm
ref}$.

A source scintillating in the regime of strong scattering may also
exhibit very fast, narrowband intensity variations due to diffractive
scintillation.  These are routinely observed in pulsars (e.g. Rickett 1970; Ewing et al. 1970).  Diffractive scintillation is only observable for source sizes $\theta_s \lesssim
r_{\rm diff} /z$.  This angular size requirement is so stringent that
no extragalactic radio source has previously been observed to exhibit
diffractive scintillation (\cite{Dennison}; \cite{Condon}).  However,
when present, it is identifiable by the fast, narrowband character of
its variations, which are distinct from the slow, broadband variations
exhibited by weak and refractive scintillation.


In this paper we present the discovery of narrowband, fast
scintillation in the quasar J1819$+$3845 at $\lambda\,21\,$cm,
characteristic of diffractive scintillation, and derive the properties
of the source component undergoing this effect.  Technical issues
related to the reduction of data from this variable source are
discussed in the following section.  In Sect.\,3 we derive the
characteristics of the variability and derive the interstellar medium
and source parameters associated with the phenomenon.  The
implications of the discovery are discussed in Sect.\,4 and the
conclusions are presented in Sect.\,5.

\section{Observations and Reductions}

Since its discovery in 1999  J1819$+$3845 has been observed 
very regularly with the 
Westerbork Synthesis Radio Telescope (WSRT). Most observations were
obtained at 4.9\,GHz but since 2001 we have increased our monitoring 
at 1.4\,GHz as well. Initially 1.4\,GHz observations were done
with the continuum backend which provides 8 contiguous bands of 10 MHz
centered around 1380 MHz. As of July 2002 we have used the new line 
backend which provides both increased spectral resolution as well as
a doubled overall bandwidth. In this paper we report on these 
wide-band observations that were taken on eleven dates between 
14 Jul 2002 and 30 Jan 2005. The last three observations were taken
with an ultra wide frequency coverage and will be described
separately below.

\subsection {Wideband (160\,MHz) data: July 2002 till Nov 2003}

All observations were continuous over a 12-hour duration. The basic
integration time of the WSRT is 10\,s but data were averaged for
either 30\,s or 60\,s.  The backend was configured to observe
simultaneously in eight 20\,MHz wide sub-bands centered at 1450, 1428,
1410, 1392, 1370, 1350, 1330/1332 and 1311/1310\,MHz.  Small gaps around
1381 and 1320 MHz were introduced to avoid occasional RFI at these
frequencies.  Each sub-band was further subdivided into 64
Hanning-tapered channels yielding 625\,kHz spectral
resolution. Further processing was performed only on the 28 odd channels
from channel 3 to 57. The lower and upper channels at the edges of each
sub-band were discarded because of higher noise levels.  Due to human
error no spectral tapering was applied to the 22 Feb 2003 data. To
retain full sensitivity channels 3-58 were processed for that epoch.
The total overall bandwidth spanned by each sub-band is
17.5\,MHz. Full polarization information was obtained but will be
presented elsewhere.  The source J1819+3845 shows only very faint ($<$
1\%) polarization which did not interfere with the total intensity
analysis.

The observations were generally taken under reasonable to fair weather
conditions (i.e. no strong winds or precipitation).  Continuous radiometry
using a frontend noise source on a 10 second time interval provided
accurate system temperatures to convert the correlation coefficients
to  relative flux density.  These relative flux densities are converted to absolute 
flux densities using the primary WSRT flux calibrator 3C286 which is tied to the Baars  
et al. (1977) scale, which assumes a flux density of 14.9\,Jy for 3C286 at a
frequency of 1380 MHz. The calibration procedure takes the spectral
index of 3C286 at these frequencies (-0.48) into account.


 \begin{figure}[h]  
 \centerline{\psfig{file=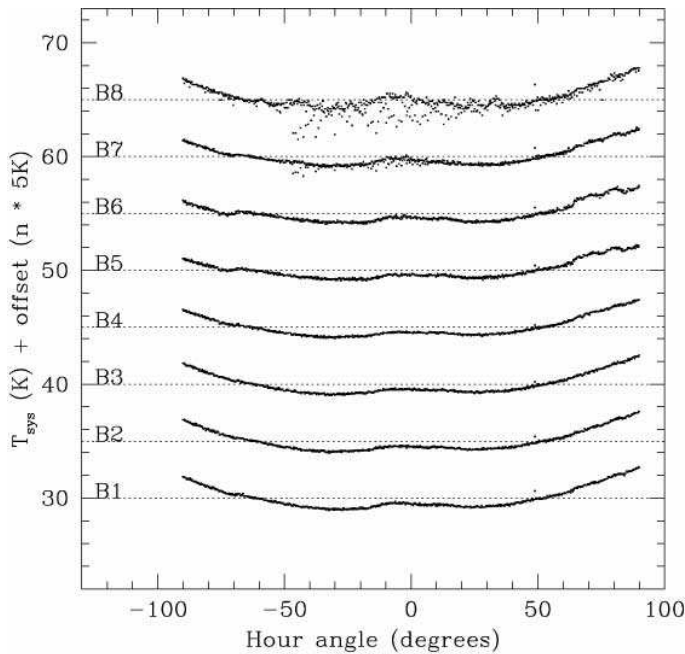,width=90mm}}
 \caption{System temperature for one telescope as a function
 of hour angle for each of the eight 20~MHz bands. Band 1 is at 1450~MHz,
 band 8 at 1310~MHz. Each curve was normalized to an average Tsys of
 about 30~K. Different bands are displaced by 5 K. Note the effects
 of the radar RFI in bands 7 and 8 (see Section 2).} \label{TsysFig} 
 \end{figure}

A typical example of the run of system temperatures
with hour angle is shown in Fig. \ref{TsysFig} for one of the 14 telescopes and all 8
frequency bands.  A slight non-linearity in the backend leads to a small underestimation
of the flux density at the extreme hour angles where the system
temperature increases slightly due to increased spillover from 
ground radiation.
We estimate this non-linearity effect to be about 1\% at most. (The
non-linearity was cured in the late spring of 2004).   Flux
density errors due to telescope pointing and telescope gain errors are
well below 1\% at 1.4 GHz.  Overall we therefore believe the flux
density scale to be good to 1-2\%. This is corroborated by the
relative flux density stability of the pairs of calibrator sources
observed before and after the 12h run on J1819$+$3845 (3C286/CTD93
before and 3C48/3C147 after). We also have long track observations on
several known stable sources which agree with our 1\% long-term
stability assessment.  

\begin{figure}[h] 
\centerline{\psfig{file=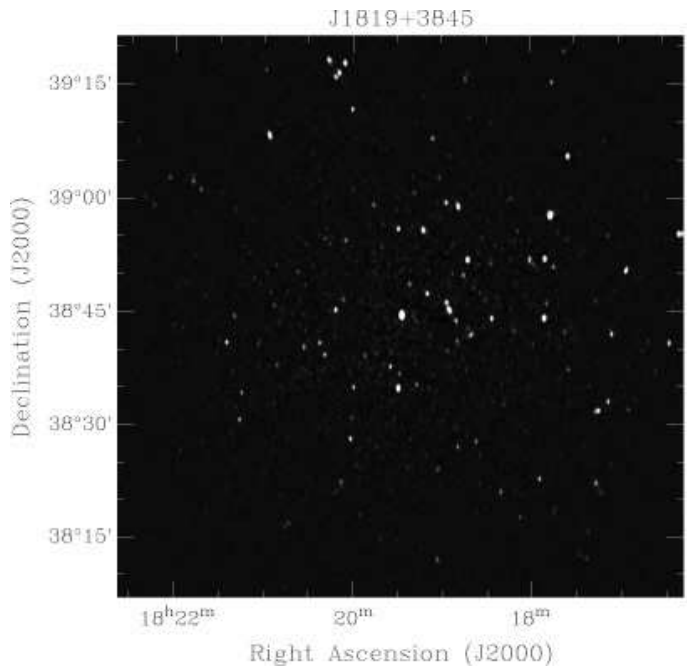,width=90mm}}
\caption{The field surrounding J1819$+$3845 as observed on 22 August 2003
at 21cm.}  \label{J1819FieldFig}
\end{figure}

The determination of weak, rapid intensity fluctuations in a
radio source is not trivial when using aperture synthesis techniques
at low frequencies, especially in an E-W synthesis array like the WSRT
where 12 hours is needed to synthesize a good beam. It would seem to 
violate the principle of `synthesis' which requires a nonvariable  sky. 
However, there are no fundamental limitations in taking care of source 
variability, as is described by Dennett-Thorpe \& de Bruyn (2000, 2003) for 
data taken on J1819$+$3845 at a frequency of 4.9 GHz.
At 1.4 GHz the situation is more complex.  Within the field of view of
the array's 25\,m dishes several hundred other sources are
detected, with flux densities, in addition to the $\sim \!\! 100$\,mJy
of J1819+3845, ranging from 13\,mJy to 0.1\,mJy.  One of the 12h images
is shown in Fig.\,\ref{J1819FieldFig}. The noise level in
the final images is typically 10-15\,$\mu$Jy per beam.

We have now observed this field about a dozen times at
1.4\,GHz. Although the character and magnitude  of the
variations of J1819$+$3845 changes significantly, the confusion from
the sky is expected, and observed, to be very stable, allowing it to
be modelled well.  Before vector averaging the visibilities for
either 30\,s or 60\,s time intervals to form the 
light curve, we removed
the response from typically 250 background sources.  
Any residual confusion from
fainter sources is estimated to be less than 1--2 mJy which is
typically 1\% of the flux density of J1819$+$3845. 
More importantly, for the present study, is that these  
residual effects are broadband in nature and would be very similar 
from epoch to epoch (because the uv-coverage is very similar).  
We are therefore certain that the observed fast and spectral
variations are not due to background confusion and must be due to the
properties of the source and the interstellar medium.

With observations spread across all seasons we have of course 
frequently observed in daytime. The quiet Sun still contributes 
a strong signal at 21\,cm despite
the $>$ 40\,dB attenuation by the primary beam. However, the visibility
function of the quiet Sun drops very fast and is  
undetectable at projected baselines beyond a few 100\,m. In a few cases 
rapid 1-2\,mJy fringing was observed and short ($\le$ 144\,m) baseline 
visibilities were excluded from the visibility averaging. 

\begin{figure*}[h] 
\begin{tabular}{rl}
\psfig{file=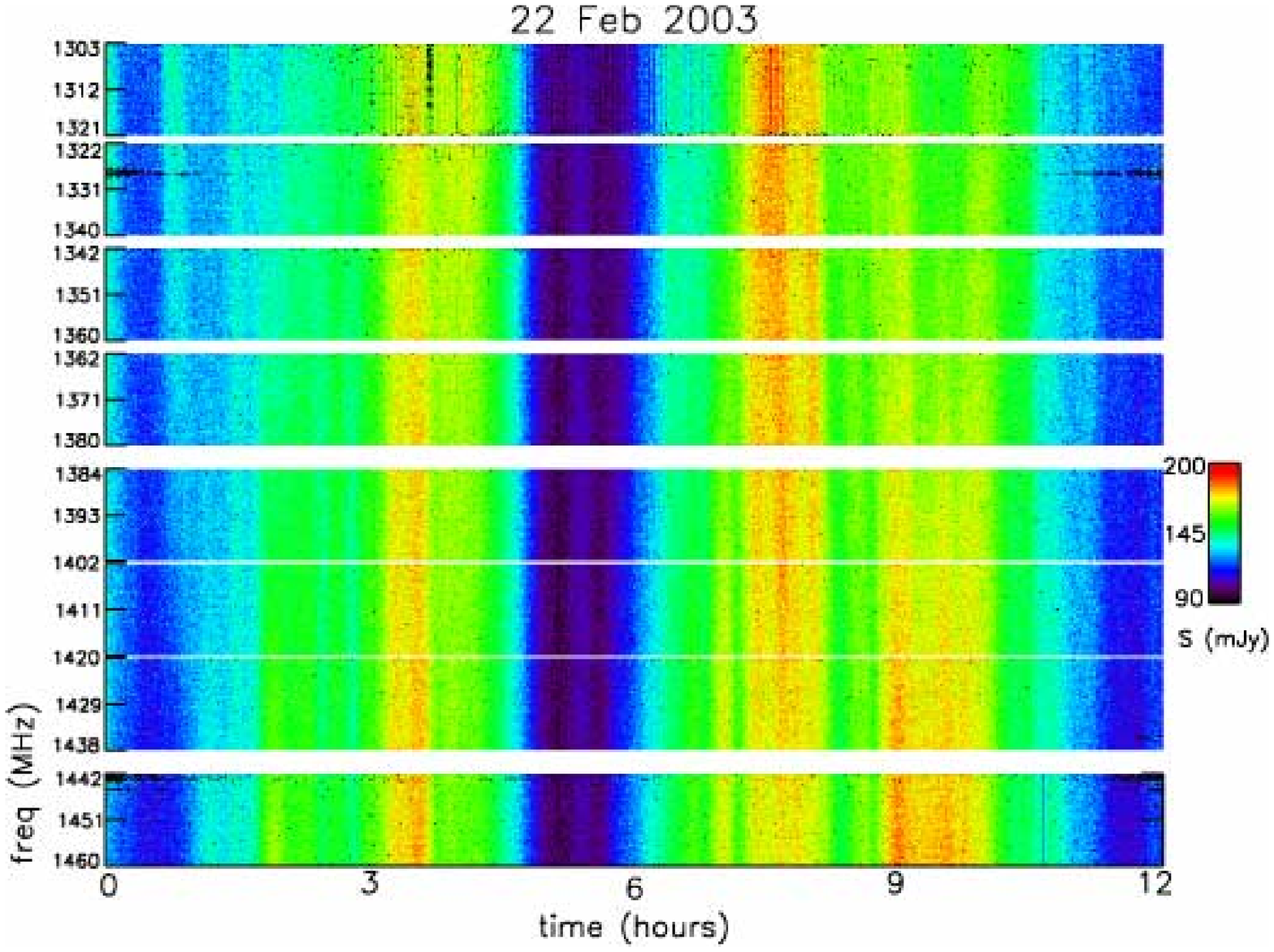,width=75mm} &
\psfig{file=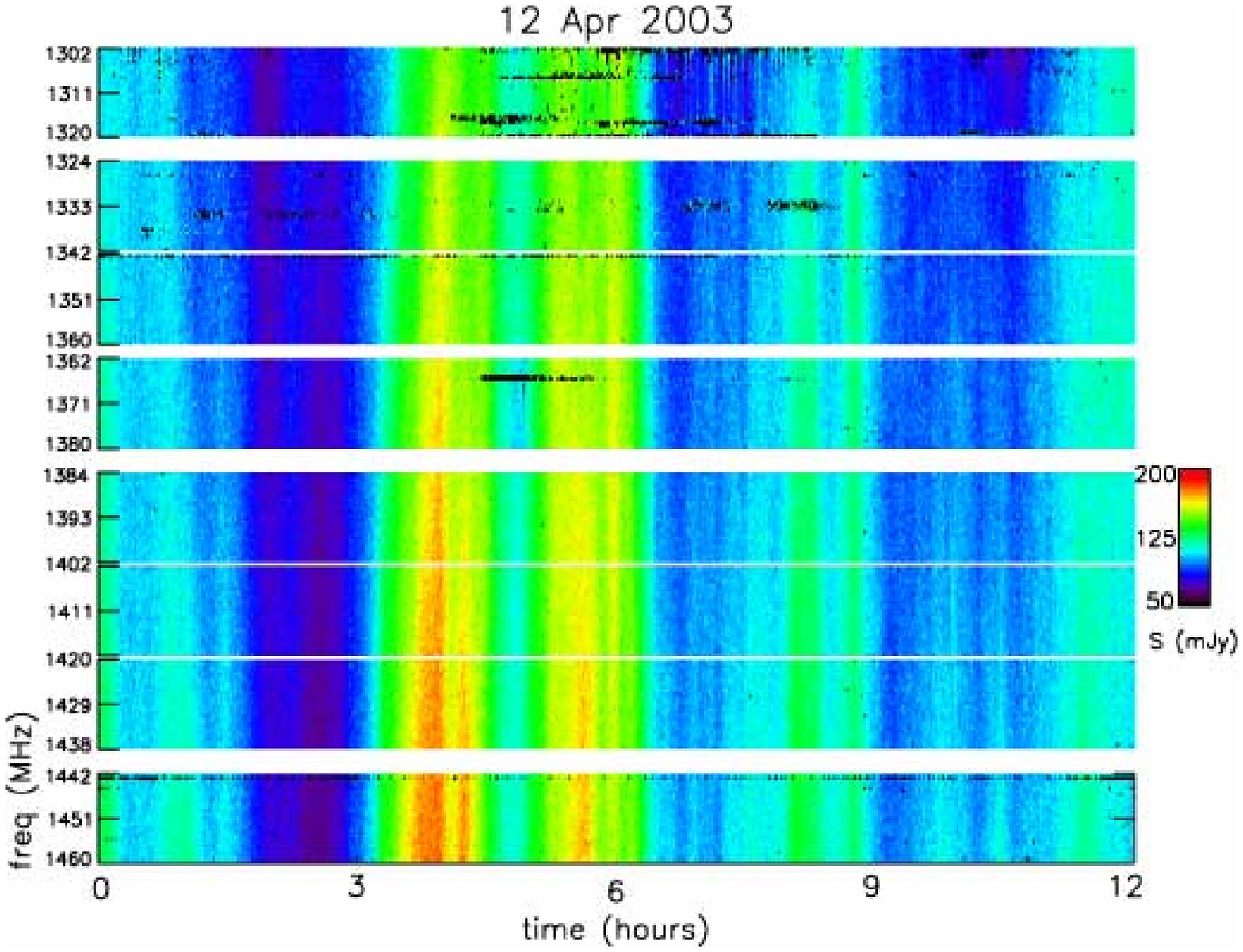,width=75mm} \\
\psfig{file=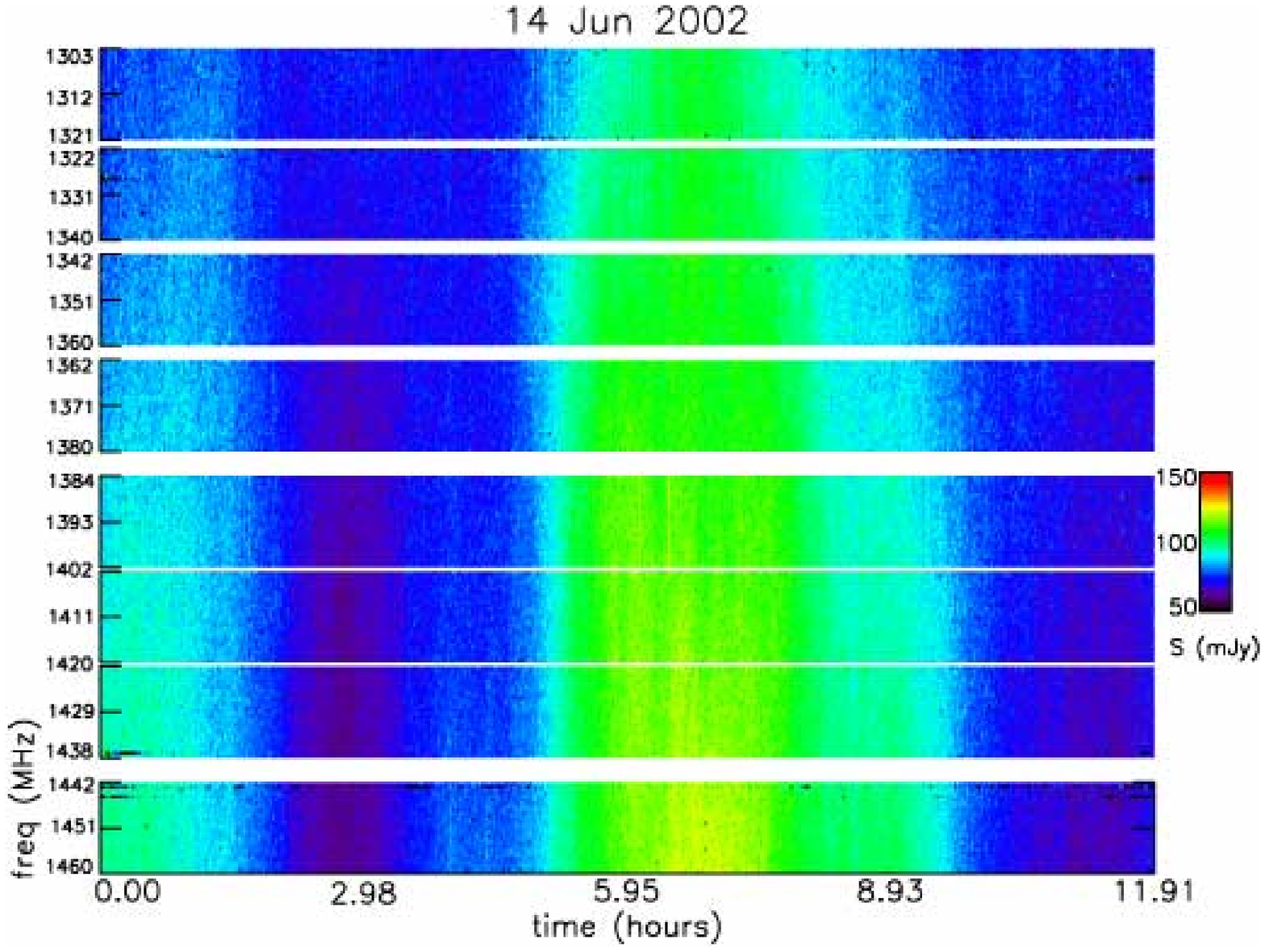,width=75mm} &
\psfig{file=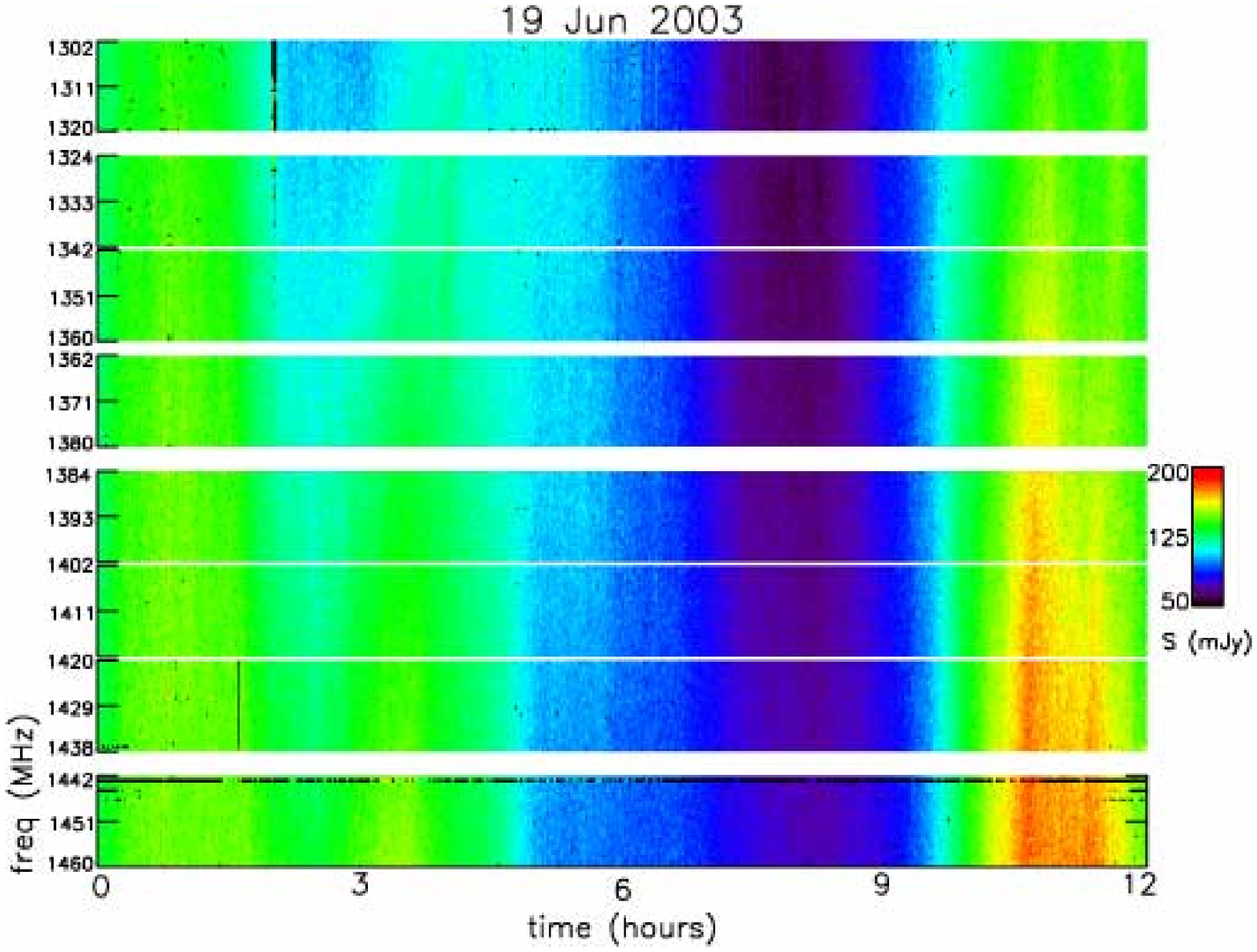,width=75mm} \\
\psfig{file=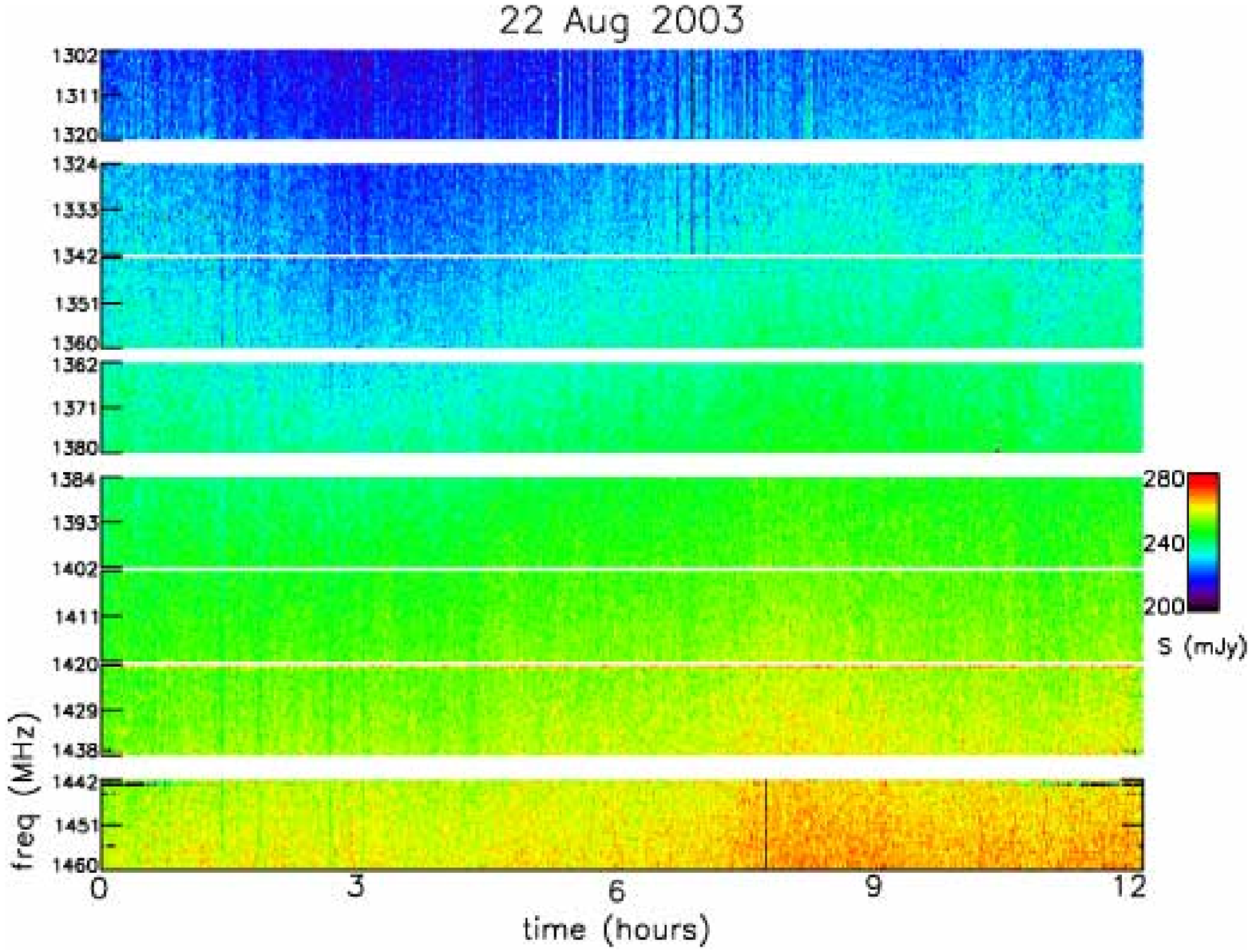,width=75mm} &
\psfig{file=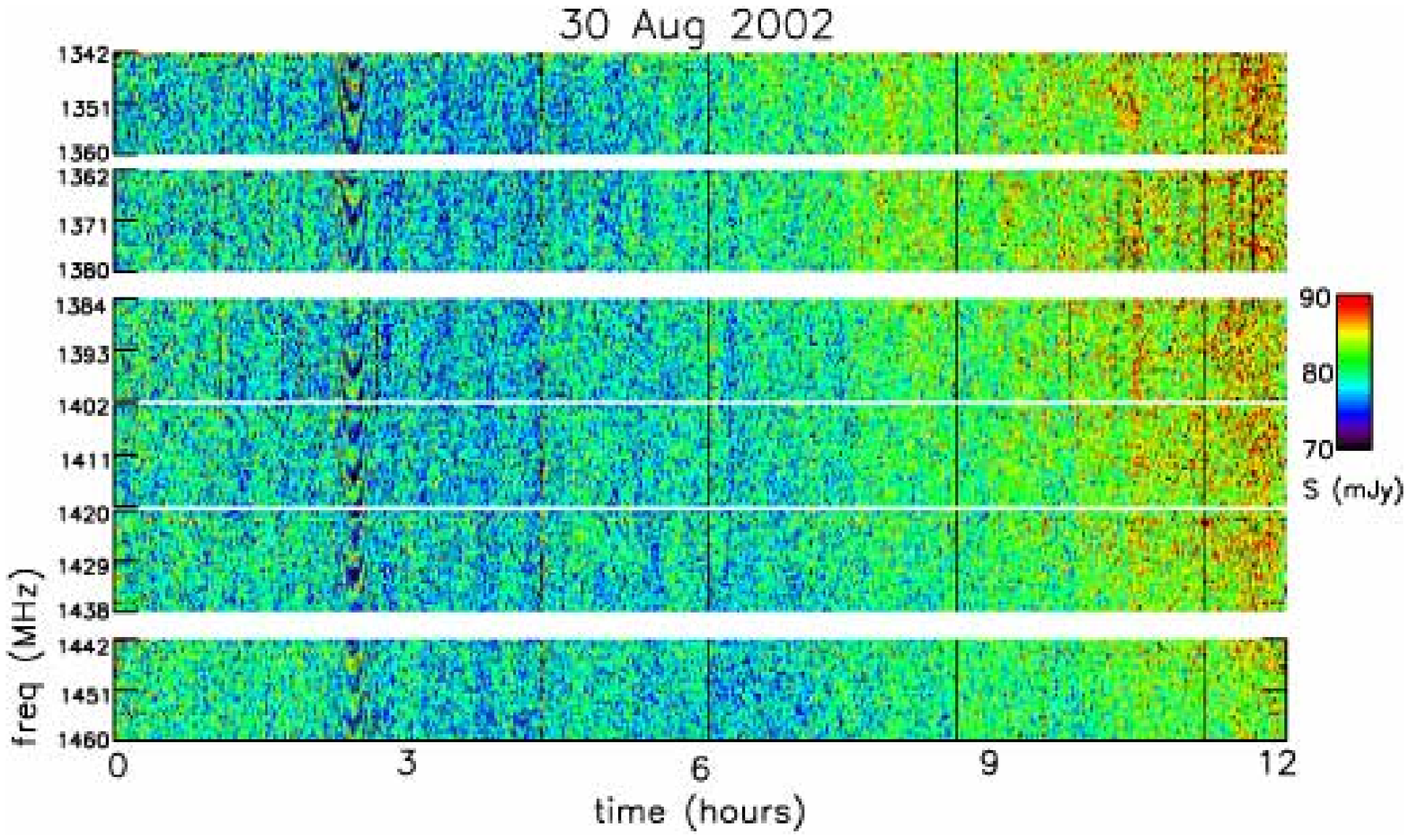,width=75mm} \\
\psfig{file=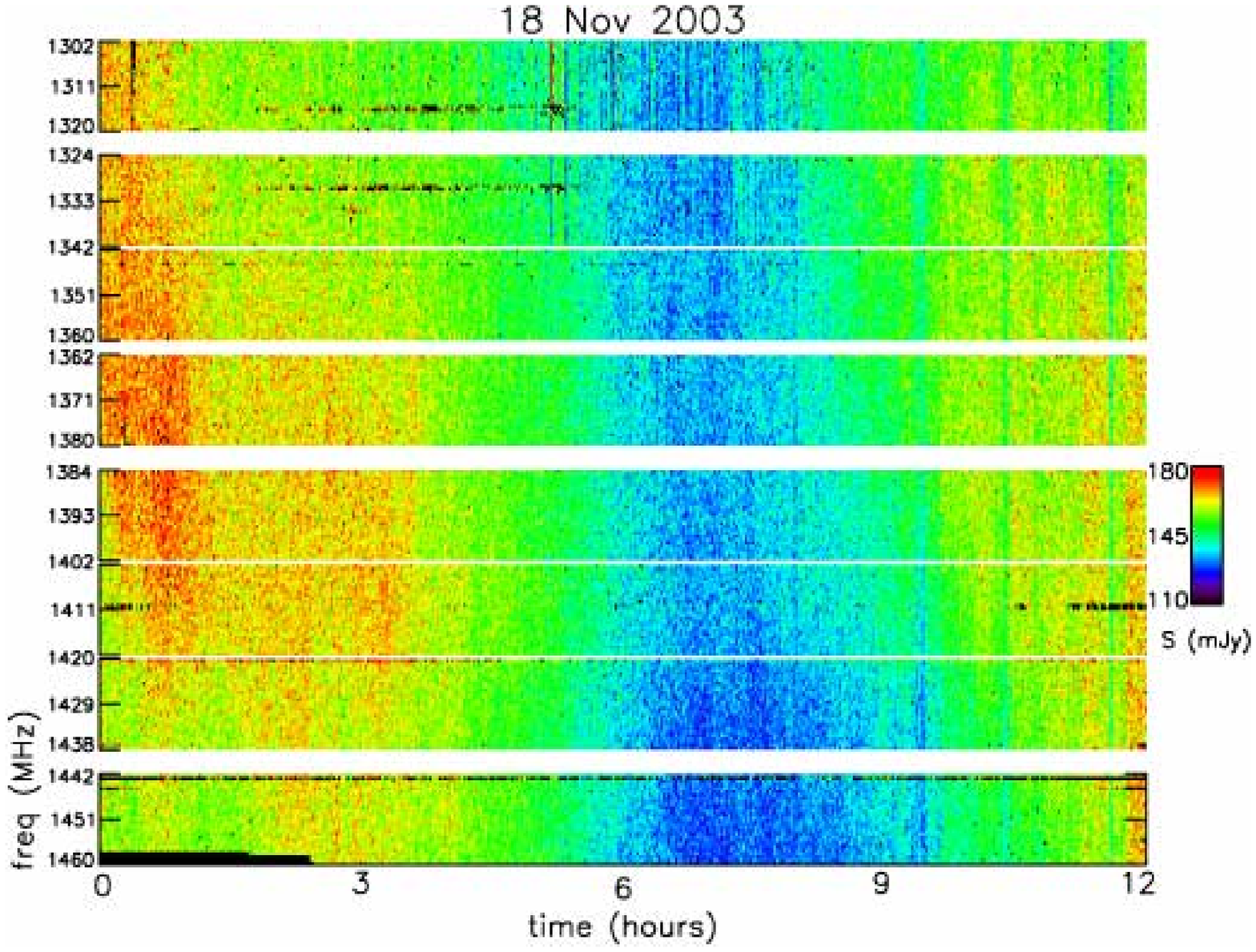,width=75mm} &
\psfig{file=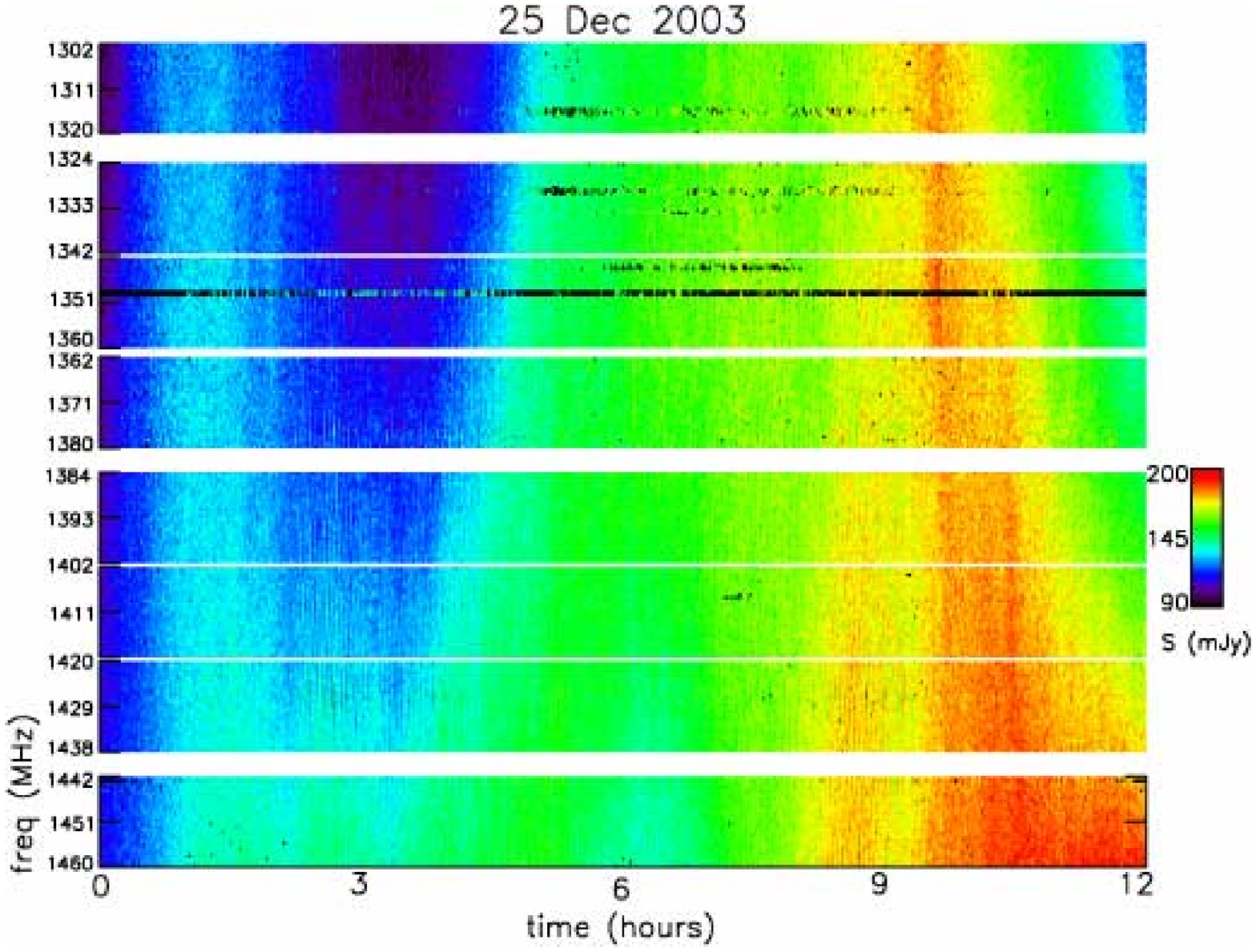,width=75mm} \\
\end{tabular} 
\caption{The dynamic spectrum of J1819$+$3845 over 160\,MHz on eight
epochs in the interval Jun 2002-Dec 2003, arranged by day of year,
given above the panels.
Each point in each dynamic spectrum represents a one minute average
over a 625\,kHz bandwidth.  Bad data flagged in the analysis are
represented by black points.  The periodic fringes visible in the
lowest frequency band on 22 Feb 2003 and 12 Apr 2003 are caused by
terrestrial interference. Note that the displayed flux density range
in some epochs when the source is not strongly variable (e.g. 30 Aug
2002) is rather small, making the dynamic spectrum appear more noisy.} 
\label{DynSpecs160MHzFig}
\end{figure*}

The lowest sub-band is intermittently affected by interference due to
1\,MHz of spectral overlap with a nearby radar.  The radar, which is
activated with a period of 9.6\,s, beats with a 4-minute period when
observed with the WSRT, whose noise sources are monitored every 10\,s.
Close inspection of the data suggests that the band centred on
1311/1310\,MHz is most affected, with weak interference also present
on the band centred at 1330\,MHz (see e.g. Fig. \ref{DynSpecs160MHzFig}, the 22 Feb 2003 dynamic 
spectrum). In the calculation of
the characteristics of the scintillation signal, data from the entire
two lower sub-bands are excluded whenever interference is evident.

Most of the RFI in the two low frequency bands enters the light curves via
the system temperature correction procedure. We have therefore also 
processed in parallel the data without applying this correction. 
The dynamic spectra for the 1310/1311\,MHz band indeed then look much cleaner. 
To take care of the slight systematic system temperature variation 
with hour angle the data for February and April 2003 shown in Fig.\,\ref{DynSpecs160MHzFig} 
were corrected using the uncontaminated Tsys curves for the higher 
frequency bands .

\begin{figure}[h]  
\centerline{\psfig{file=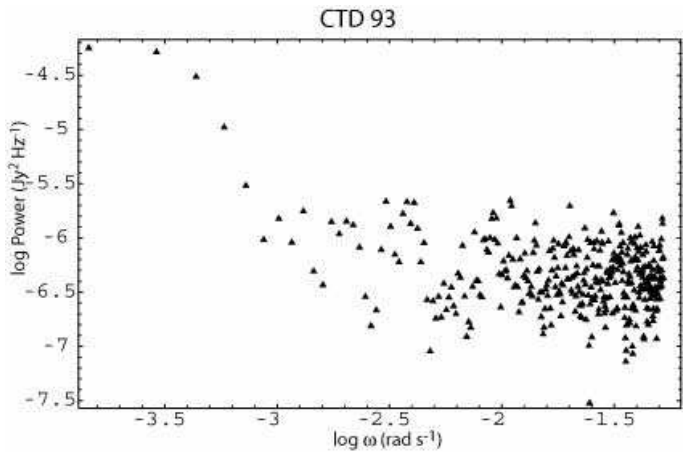,width=90mm}} 
\caption{The power spectrum of temporal variations from a 10-hour
observation of the stable bright radio source CTD~93.  In order to
compare with the temporal variations in J1819$+$3845, the power
spectrum is computed from a light curve in which the flux densities
have been reduced (by a factor $\sim \!\! 50$) to have a mean of
100\,mJy (see \S2).} \label{CTD93PowerFig}
\end{figure}

In order to ascertain the accuracy of the overall amplitude
calibration on a range of timescales, we have also reduced a 10 hour
observation of the bright stable radio source CTD93, observed in May
2003 with a similar instrumental setup as J1819$+$3845.  The power
spectrum of the temporal fluctuations of this source is shown in
Fig.\,\ref{CTD93PowerFig} after scaling the intensity to a mean flux of 100 mJy.  This
means that the thermal noise has been reduced to an insignificant level
and we are left with the combined variations due to atmospheric opacity,
pointing and amplitude calibration. The level of `variability'
observed in CTD93 on frequencies of $<$ 0.001 rad/s, 
which correspond to timescales of about 10m to 2h, is significantly less than
1\%.  On faster timescales this drops to about 0.1\%, a level
probably set by the total power amplitude calibration. All
fluctuations observed in J1819$+$3845 appear to be significantly 
in excess of these levels, at any temporal scale, and become 
thermal noise limited at the fastest timescales sampled.


\subsection {Ultra wideband ($\sim$600\,MHz) data: Jan 2004 to Jan 2005}
 
As of Jan 2004 the WSRT online software allowed frequency switching with
high efficiency between different frequency settings within the L-band
receiver which covers the range from 1150 -- 1800 MHz. Although a
significant fraction of this band is affected by man-made interference
-- GPS, GLONASS and geostationary satellites -- the advantage of the
wider band to study the spectral decorrelation effects of the
scintillations more than outweighed this loss of data.  Three
observations were taken with this ultra wideband setup, which had the
following switching scheme: at intervals of 60 seconds we switched
between different frequency `combs' of 8 adjacent 20~MHz bands.  For
every comb the first 20\,s of data had to be discarded leaving 40\,s of
good data.  The observations of 25 Jan 2004 and 12 April 2004 were carried
out with three frequency combs but different central frequencies.  In
the most recent data of 30 Jan 2005 we used 4 combs almost completely
covering the available L-band frequency range.  After calibration and
RFI editing the data were averaged over 40\,s timeslots leading
to light curves sampled on a regular 180\,s or 240\,s grid.

The strong RFI encountered in several bands of each comb made it
impossible to provide a reliable intensity calibration.  The
decomposition of the total power radiometric data into system
temperatures and electronic gains requires stable conditions during a
10\,s period, which is obviously not the case under strong and
impulsive RFI conditions. The effects of this on the amplitude
stability were exacerbated by the small linearity problem in the
receiver.  The ultra-wideband data were therefore internally
calibrated on a band-by-band basis for each band of the
frequency combs by normalizing on the 12-h averaged flux of J1819$+$3845
itself.

 
\section{Results and Analysis}

\begin{figure*}[h] 
\vskip 1cm \centerline{\psfig{file=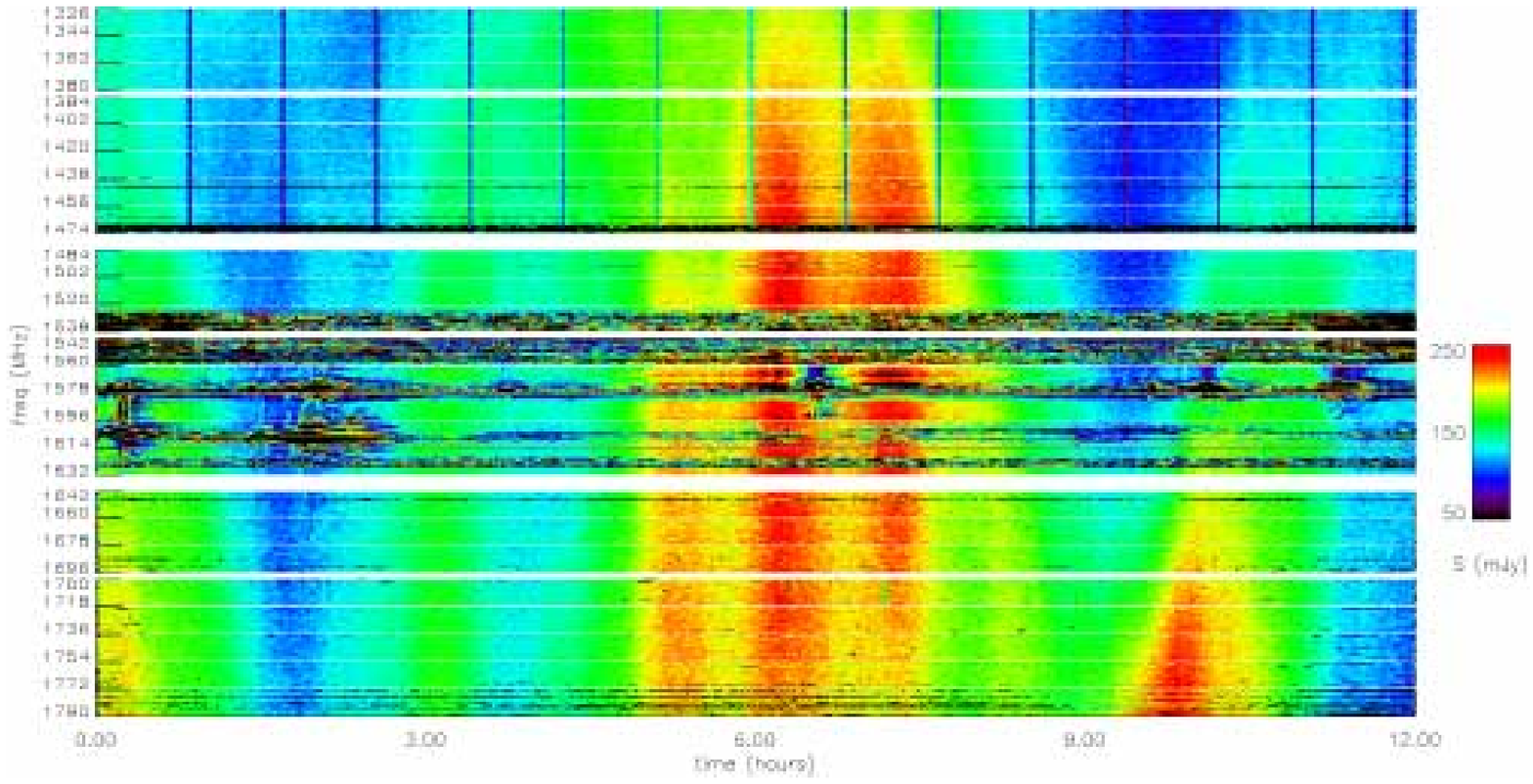,width=110mm}
}
\vskip 1cm \centerline{\psfig{file=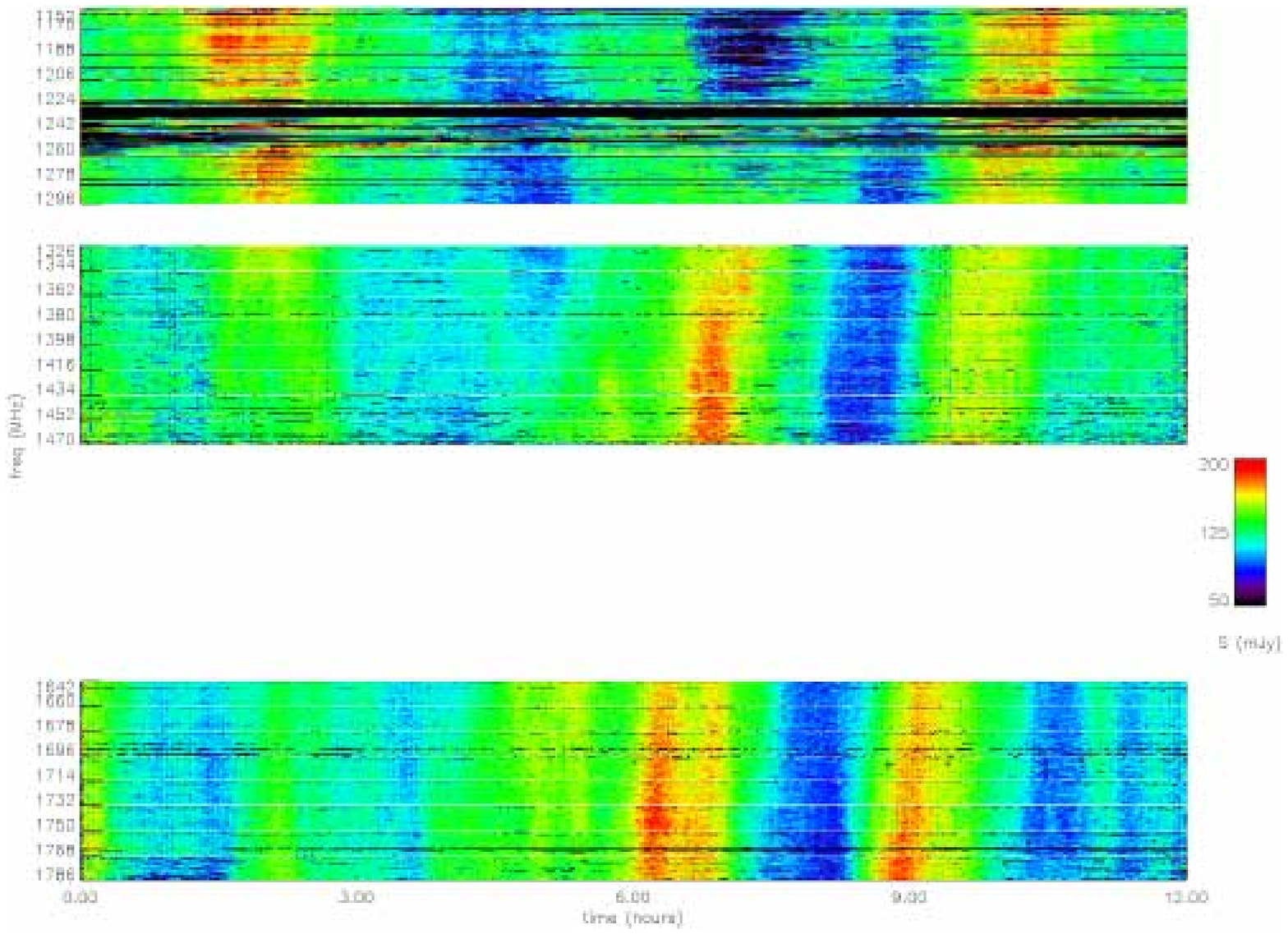,width=110mm}
}
\centerline{\psfig{file=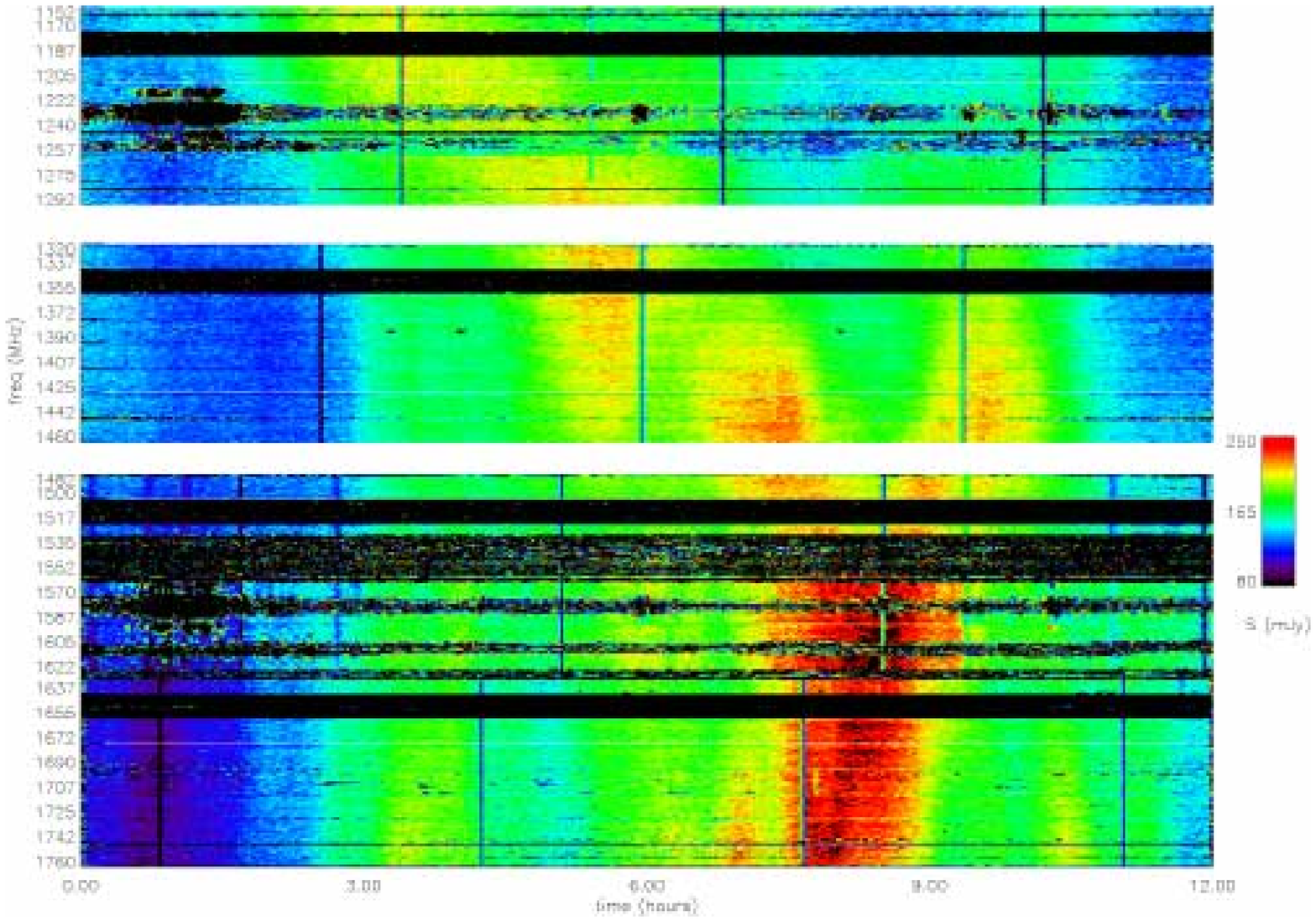,width=110mm} }
\caption{Wideband, frequency-mosaiced observations of the
scintillation spectrum in J1819$+$3845 on 25 Jan 2004 (top), 12 Apr
2004 (middle) and 30 Jan 2005.  The sizes of the white vertical spaces
between data channels are proportional to their corresponding
frequency gaps.  Bad data points, channels or bands are marked by 
black pixels.  The contribution of the intrinsic spectrum of J\,1819$+$3845
has been removed by normalising each spectral channel so that they are
identical (and equal to the mean source flux density, averaged over
the entire dynamic spectrum).} 
\label{WideBandDynSpecsFig}
\end{figure*}

The dynamic spectra displayed in Fig. \ref{DynSpecs160MHzFig} and Fig.
\ref{WideBandDynSpecsFig} present a concise
summary of the intensity fluctuations exhibited by J1819$+$3845 over
all eleven epochs of our observations.  The variations exhibit fine
structure in both time and frequency.  The spectral features are
stochastic in nature, as both fading and brightening streaks are
visible in all dynamic spectra.  The spectral structure is associated
with the fastest variations visible during each epoch.  This is
particularly apparent during the 22 Feb 2003 and 12 Apr 2003 observations,
in which variations occur on timescales as short as 20\,minutes, but
are as long as several hours at other times of the year, with a
variation being defined here as a complete oscillation in the
light curve.  The reduced duty cycle (40\,s of data for every 
180\,s or 240\,s)  in the frequency-mosaiced dynamic 
spectra in Fig.\,\ref{WideBandDynSpecsFig} means that some 
of the fine temporal structure evident in other 
observations (c.f. 22 Feb 2003, 12 Apr 2003) would not be as easily 
detectable in these observations. A very recent regular 160\,MHz 
observation (taken on 28 Mar 2005), not presented here shows
that such fine structure is still present. A more detailed analysis of 
variations in the 21\,cm band over a period of 6 years will
be presented in de Bruyn et al. (in preparation).


The light curves shown in Fig.\,\ref{LightcurvesFig} further indicate that the intensity
variations occur on several timescales and that these timescales
change as a function of observing epoch.  This is also demonstrated by
the power spectra of the intensity fluctuations shown in Fig.\,\ref{PowerSpectraFig}.  The light curves from which these power spectra were computed contained no gaps over the 12 hour duration of the observation, with flux densities sampled every 60\,s.

We discuss the temporal variability in Sect. \ref{TemporalData} and
the spectral characteristics in Sect.\,\ref{Spectral}.  These are used
to derive parameters of the source and scattering medium in
Sect.\,\ref{Fit}.

\subsection{Timescales} \label{TemporalData}

Here we compare the observed power spectrum of temporal intensity
variations to models for refractive scintillation.  We argue that
refractive scintillation fails to account for much of the variation
observed on timescales shorter than $6\,$hours.

\begin{figure*}[h] 
\centerline{\psfig{file=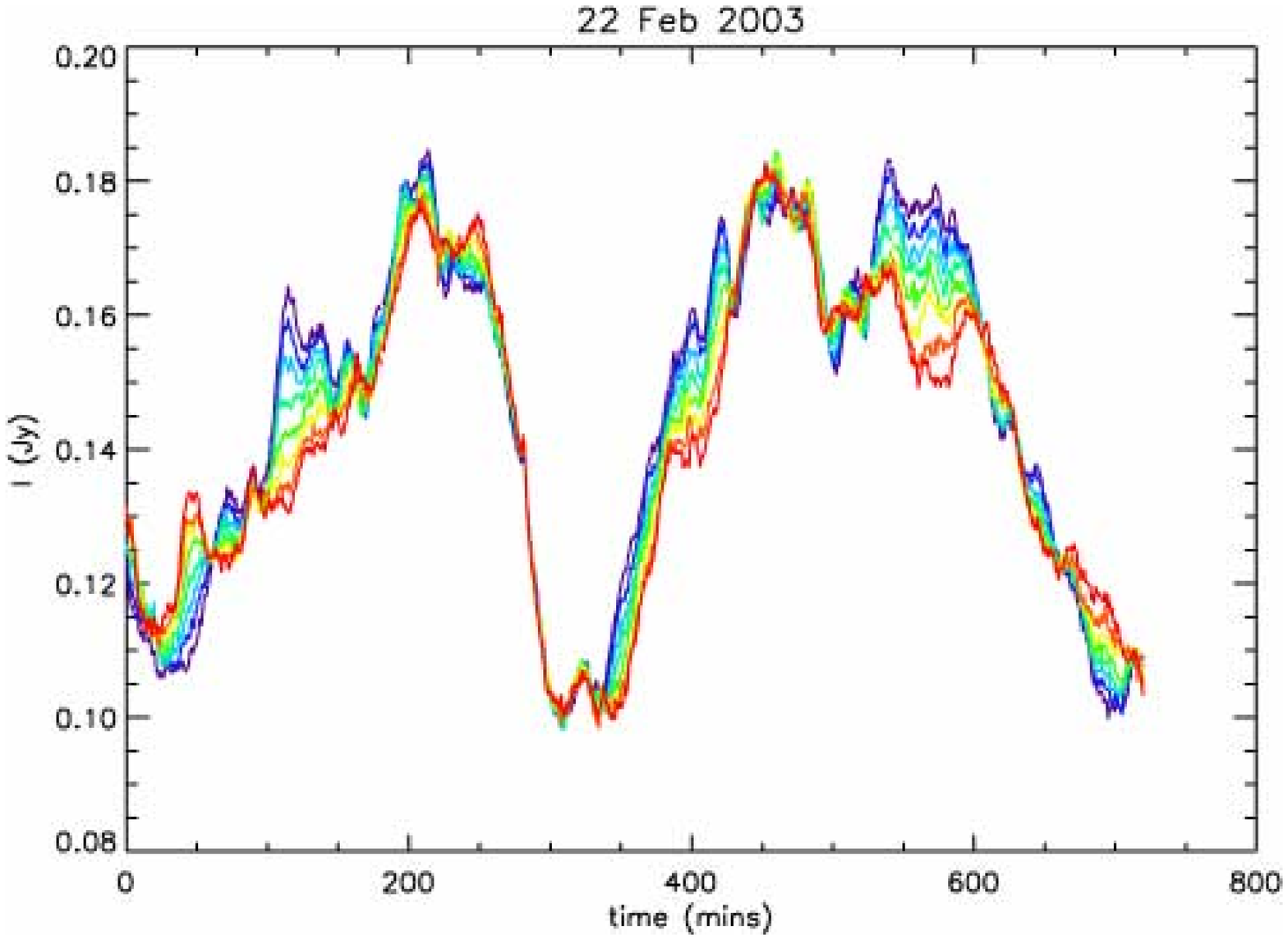,width=100mm} }
\centerline{\psfig{file=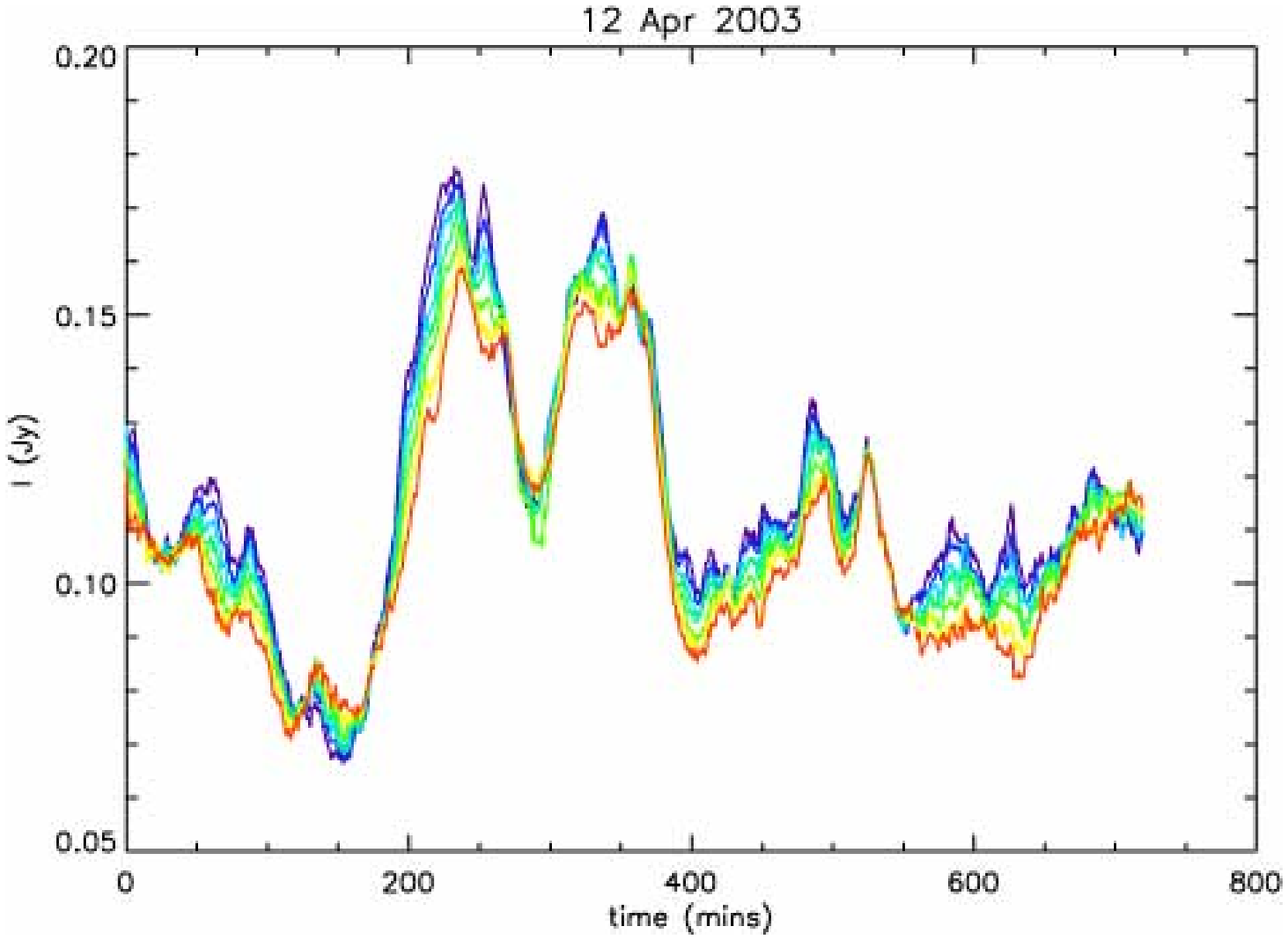,width=100mm} }
\centerline{\psfig{file=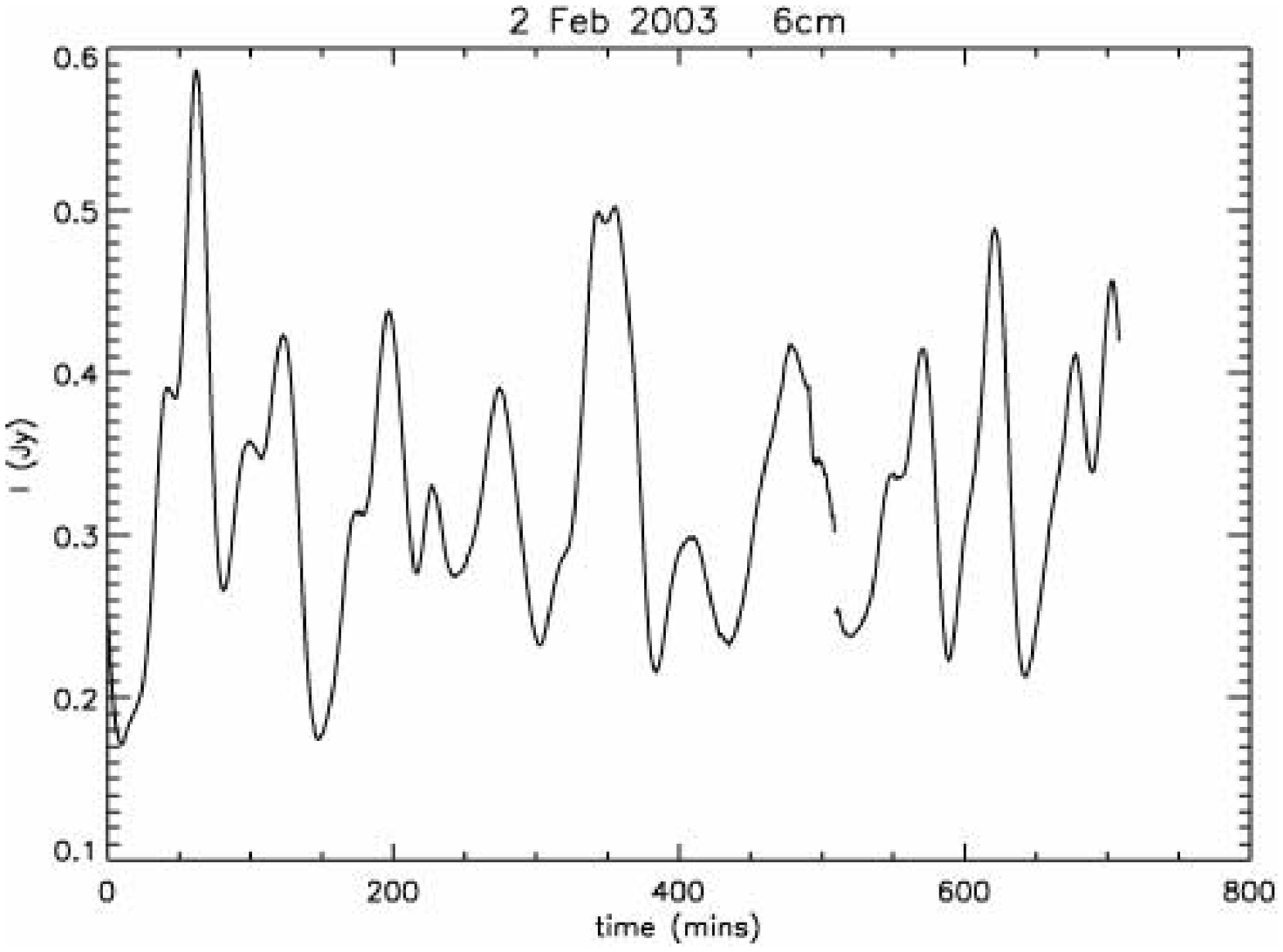,width=100mm} }
\caption{Light curves of the variability on 22 Feb 2003 and 12 Apr 2003
at 1.4\,GHz in each band.  The (colour-coded) light curves from
individual bands differ on timescales ranging from 20 mins up to
several hours.  The variations at the lowest frequency band in the 12
Apr 2003 dataset were contaminated by terrestrial interference, and
are excluded from this plot.  The observing frequency increases with
colour from red to blue.  The differences between the band light curves
are well above the 1\,mJy noise level (per 60~s).  
The bottom panel, a light curve
of the source at 4.8\,GHz on 2 Feb 2003, contrasts the smoothness of
the intensity variations observed in the regime of weak scattering
with those observed at 21\,cm.  The noise on the 4.8\,GHz light curve is 
typically 1\% of the flux density or less.  These errors are discussed in Dennett-Thorpe \& de Bruyn (2003) in detail. } \label{LightcurvesFig}
\end{figure*}

The scintillation velocity is fastest in the period December-April
(Dennett-Thorpe \& de Bruyn 2003), so the 22 Feb 2003 and 12 Apr 2003
datasets, which exhibit the fastest intensity variations, are the most
useful in understanding the temporal characteristics of the
variations.  Fig. \ref{DynSpecs160MHzFig} and Fig. \ref{LightcurvesFig} clearly demonstrate that, for both
datasets, narrowband ($\sim 160\,$MHz), $\sim\!\! 20-120$-minute variations
are superposed on slower $\sim\!\! 6$-hourly variations.  (The
frequency-mosaiced observations made on similar dates in 2004 and 2005
are less suitable because their time-sampling is irregular, and their
decreased S/N (per frequency channel) renders them unsuitable for
characterising any fast, low-amplitude intensity fluctuations.)

The variations on timescales $\ga 6\,$hours match those expected from
refractive scintillation on the basis of observations at higher
frequencies.  The transition between and strong scattering is thought
to occur in the range $\ga \!\!3.5 - 5\,$GHz for this source
(\cite{DennettThorpe00}) and the intensity variations observed at
4.9\,GHz are attributed to a scattering screen $\sim 15\,$pc from
Earth, with $\viss \approx 50$\,km\,s$^{-1}$.  On this basis and assuming Kolmogorov turbulence one
predicts the refractive scintillation timescale at 1.4\,GHz to be
between 4 and $8\,$hours, consistent with the slow variations observed
here.

The larger duration of the frequency-dependent scintles observed
during other epochs, their lack of associated large broadband flux
density deviations and the increased timescale of the fine,
frequency-dependent structure suggests that the refractive
scintillation timescale exceeds the 12-hour span of these
observations.  This is expected on the basis of the slow-down observed
in the variations at 6\,cm during this period.

Fig.\,\ref{LightcurvesFig} displays the variations seen at 1.4\,GHz with those observed at
4.9\,GHz during the same period.  This demonstrates an excess of
variability on short (20-120 minute) timescales at 1.4\,GHz.  It is
difficult to account for this excess in terms of refractive
scintillation alone, since the timescale of the variations increases
sharply ( $\propto \lambda^{2.2}$ (e.g. Narayan 1992; Armstrong, Rickett \& Spangler 1995)) in the regime of strong scattering.

To further illustrate the difficulty of accounting for all of the power observed in the short timescale variations in terms of refractive scintillation,  we consider two quantitative models for the power spectrum of refractive variability: scintillation (i) from a thin phase-changing screen and (ii) from an extended medium in which the source size exceeds the refractive scale (i.e. $\theta_{\rm src} > \theta_{\rm ref}$).

The temporal power spectrum due to refractive scintillation caused by a thin screen of scattering material is (\cite{Codona}) 
\begin{eqnarray}
\Phi_I (\omega) &=& \frac{4\, r_e^2 \, \lambda^2 I_{\rm src}^2 \, \Delta L}{\viss} \int d \kappa_y \Phi_{N_e} \left( \frac{\omega}{\viss},\kappa_y \right) \left\vert V \left( \frac{\omega \, z}{\viss \, k},\frac{ \kappa_y z}{k} \right) \right\vert^2 \nonumber \\ &\null&
\quad \times \sin^2 \left[ \frac{ \left( \frac{\omega^2}{\viss^2}+\kappa_y^2 \right) z}{2 k} \right] \exp \left[ - D_\phi \left( \frac{\omega \, z}{\viss \, k},\frac{\kappa_y z}{ k} \right) \right], \label{TempRefPow}
\end{eqnarray}
where $\omega$ is the angular frequency, $\kappa_x$ and $\kappa_y$ are spatial wavenumbers, $\Phi_{N_e}(\kappa_x,\kappa_y)$ is the power spectrum of electron density fluctuations, $z$ is the distance to the scattering medium, $\Delta L \ll z$ is the screen thickness, $r_e$ is the classical electron radius, $V({\bf r})$ is the visibility of the source and $D_\phi({\bf r})$ is the phase structure function.  Only the source visibility can counteract the sharp decline in the power spectrum due to the exponential function at $\omega > \viss/r_{\rm ref}$, but this requires a visibility that {\it rises} nearly exponentially quickly to account for the observed shallow of the decline of power spectrum (see Fig.\,\ref{PowerSpectraFig}).  

The refractive power spectrum for a  scattering medium distributed along the line of sight declines more slowly at high temporal frequencies.  For a source of angular size $\theta_{\rm src} \ga \theta_{\rm ref}$ scattered in a medium with thickness $L$ and distributed according to $C_N^2(z)=C_N^2(0) \exp(-z^2/L^2)$ (with $z$ measured from the observer) the power spectrum of intensity fluctuations is (\cite{Coles})
\begin{eqnarray}
\phi_I(\omega) &=&  
\frac{2 \sqrt{\pi} \, r_e^2 \lambda^2 L\, I_{\rm src}^2 }{\viss} \int d \kappa_y  \Phi_{N_e} \left( \frac{\omega}{\viss},\kappa_y \right) \, \nonumber \\
&\null& \qquad \times \frac{1-\exp \left[ - 
\frac{(\omega^2/\viss^2 + \kappa_y^2)^2 r_{\rm F}^4}{ 4[1+ (\omega^2/\viss^2 + \kappa_y^2) L^2 \theta_s^2/2]} \right] }
{\sqrt{1+\left( \frac{\omega^2}{\viss^2} + \kappa_y^2 \right) L^2 \theta_s^2/2}}. \label{ExtTempRefPow}
\end{eqnarray}
For a Kolmogorov spectrum of turbulent fluctuations this power spectrum declines asymptotically as $\sim (v_{\rm ISS}/\omega)^{8/3}$. 

Fig.\,\ref{PowerSpectraFig} illustrates the excess of power observed on short timescales
relative to that expected from the two refractive scintillation models
discussed here.  The fitted models assume a scintillation speed
$\viss=50\,$km\,s$^{-1}$ and that all of the source emission is compact
enough to be subject to interstellar scintillation (i.e. there are no
$\ga 5 \,$mas features in the source).  The fit parameters are listed in Table 1.  
The level of fluctuations on timescales of less than $\sim 2\,$hours, which the dynamic spectra show
to be associated with highly frequency dependent variations, are difficult to
explain in terms of refractive scintillation, suggesting that their
origin is most likely diffractive in nature.

\begin{table*} 
\begin{center}
\begin{tabular}{|c|c|c|}
\hline
\null & 22 Feb 2003 & 12 Apr 2003  \\ \hline
$z=15\,$pc thin screen matching lowest frequency power & $C_N^2 L = 1.82 \times 10^{16} $\,m$^{-17/3}$ & 
$C_N^2 L = 1.38 \times 10^{16} $\,m$^{-17/3}$  \\
$z=4\,$pc thin screen matching lowest frequency power &$C_N^2 L = 1.42 \times 10^{17} $\,m$^{-17/3}$ & 
$C_N^2 L = 1.01 \times 10^{17} $\,m$^{-17/3}$  \\  
$z=15\,$pc thin screen matching second lowest frequency power & $C_N^2 L = 3.69 \times 10^{15} $\,m$^{-17/3}$  & $C_N^2 L = 4.0 \times 10^{15} $\,m$^{-17/3}$ \\ 
extended medium ($\Delta L = 15\,$pc) & $C_N^2 = 10^{-3.18} $\,m$^{-20/3}$ &  $C_N^2 = 10^{-3.25} $\,m$^{-20/3}$ \\ \hline 
\end{tabular}
\end{center}
\caption{Parameters used in the fits to the temporal power spectra in Fig.\,\ref{PowerSpectraFig}. The transition frequency between weak and strong scattering for $C_N^2 L = 1.42 \times 10^{17}$ at $z=4\,$pc is 5.45\,GHz and for $z=15$\,pc $C_N^2 L$ values of $1.82 \times 10^{16}$ and $3.69 \times 10^{15}\,$m$^{-17/3}$ correspond to transition frequencies of 3.89 and 2.22\,GHz respectively.}
\end{table*}

\subsection{Spectral decorrelation of the scintillation signal} \label{Spectral}

The spectral characteristics of the scintillation signal are
determined by computing the autocovariance of the intensity
variations across the observing band, $C_\nu(\Delta \nu) = \langle
I(\nu+\Delta \nu) I (\nu) \rangle - \bar{I}^2$.

In order to isolate the spectral decorrelation due to the narrowband
fluctuations only, it is necessary to remove the appreciable variation
in the mean (intrinsic) source flux density across our observing
bandwidth.  This is removed by `flat-fielding' the spectrum prior to
autocorrelation.  The dynamic spectrum is normalised so that the
time-average flux density in each spectral channel is identical, and
equal to the flux density averaged across the entire dynamic spectrum.
This prevents the mean spectral slope across the band from
masquerading as a scintillation signal and weights intensity
fluctuations across the entire spectrum fairly, provided that the
spectrum of the narrowband variations resembles the overall source
spectrum.  Although this spectral match may only be approximate in
practice, the normalisation is sufficient for present purposes since
the contribution to the error caused by any small spectral index
mismatch is also small\footnote{The error is no larger than the mean
square of the flux density error; the contribution cancels to first
order in flux density difference since the autocorrelation averages
out the contributions from both higher and lower frequencies.  For
instance, the additional contribution to the autocorrelation is $\sim
0.1$\,mJy$^2$ across 100\,MHz for a diffractively scintillating
component of spectral index 1.0 and a source with spectral index
0.3.}.  The effects of spectral misestimation are incorporated in the
error budget when considering fits to the autocorrelation functions in
\S\ref{Fit} below.
 
The short timescale of the scintillation requires that $C_\nu$ be
computed for each one-minute time slot individually.  The final
frequency autocorrelation function is the average of the $710-720$
functions computed separately from each time slot.

The autocovariances for the 25 Jan 2004, 12 Apr 2004 and 30 Jan 2005
observations are shown in Fig.\,\ref{ACFFigs}.  These datasets are used because they 
span a sufficiently large spectral range that they encompass the bandwidth of the scintillation
structure.  The error in our sample autocorrelation function, depicted
by the grey region in Fig.\,\ref{ACFFigs}, incorporates the fact that our
observations sample a finite number of scintles both in time and
frequency (see Appendix \ref{AppErrs}).


Can refractive scintillation alone account for the spectral structure?
No calculation of the form of the spectral decorrelation due to refractive scintillation exists in the literature, but refractive scintillation is known to decorrelate on a bandwidth comparable to the
observing frequency: $\Delta \nu_{\rm ref} \sim \nu$  (e.g. Goodman \& Narayan 1989).   We can make a simple estimate of the importance of refractive intensity variations by following the simple geometric optics model described in Narayan (1992).  The amplitude of a typical flux variation depends on the root-mean-square focal length of  phase fluctuations on scales $r_{\rm ref}$ in the scattering medium.  Since the focal length is much larger than the distance to the observer, and assuming Kolmogorov turbulence, the amplitude is $\sim r_{\rm diff}^{1/3} (z/k)^{-1/6}$.  Thus the expected flux density change across a
bandwidth $\Delta \nu$ is $\Delta S (1+\Delta \nu/\nu)^{17/30}-\Delta
S$, where $\Delta S \approx 25 \,$mJy is the
root-mean square amplitude of the refractive scintillations.  Thus the
contribution due to refractive frequency variations is at most
$2.0\,$mJy across a bandwidth of $200$\,MHz, considerably smaller than
the variation observed.

Larger chromatic effects are possible if a strongly refracting `wedge'
or prism of interstellar material is present along the line of sight.
Chromatic refraction due to a wedge would be important if its gradient
were sufficient to change the angle of refraction from the low to high
edge of the observing band by an amount comparable to $\theta_{\rm
ref}$.  However, if the gradient is not perfectly aligned orthogonal
to the direction of the scintillation velocity one expects the
scintillations to be displaced in time as well as frequency.  The
wedge would have to displace the scintillation pattern by a relative
distance $\ga 0.3 \, r_{\rm ref}$ from one edge of the observing band
to the other to account for the chromatic nature of the intensity
fluctuations observed here.  One would then expect the wedge to also
displace the scintillations in time by $\ga 0.3\, t_{\rm ref}$ as one
moves across the observing band.  No such temporal displacement is
observed.  Moreover, as the scintillation velocity changes through the
year, one would expect a systematic change in the slope of the
frequency-dependent scintles (i.e. $d\nu/dt$) with a change in the
direction of the scintillation velocity.  As no systematic change with
scintillation velocity is observed, we conclude that there is no
spectral contamination due to a refracting wedge.

\subsection{Diffractive scintillation source characteristics} \label{Fit}

The temporal and spectral characteristics of the intensity variations
are combined to determine the parameters of the source undergoing
diffractive scintillation.  We concentrate on fits to the spectral
decorrelation of the 25 Jan 2004, 12 Apr 2004 and 30 Jan 2005 datasets
where the spectral coverage exceeds the typical bandwidth of the
scintillation structure.

A detailed interpretation of the scintillation parameters depends in
detail on whether the scattering material is located in a thin layer
or whether it is extended along the ray path.  The manner in which the
scale size of the scintillation pattern and the form of the spectral
decorrelation are altered by source size effects depends on the
distribution of scattering material along the line of sight.  We
consider two specific models, one in which the scattering material is
confined to a thin screen a distance $z$ from the observer, and the
other in which the scattering material is extended homogeneously out
to a distance $\Delta z$ from the observer.

In the thin screen model the spectral autocorrelation takes the form (\cite{Chashei}; \cite{Gwinn}) 
\begin{eqnarray}
F_{\rm thin} = A_{\rm off} + S_{\rm diff}^2 \left( 1 + \frac{\theta_0^2}{\theta_{\rm cr}^2} + \frac{\Delta \nu^2}{\Delta \nu_{\rm t}^2} \right)^{-1}, \label{ACFthin}
\end{eqnarray}
where $S_{\rm diff}$ is the flux density of the source component exhibiting diffractive scintillation, $\Delta \nu_{\rm t} =  \nu r_{\rm diff}^2/r_{\rm F}^2$ is the decorrelation bandwidth that a {\it point source} would
possess in this scattering medium, and $\theta_0/\theta_{\rm cr}$ is the ratio of the source angular radius to a critical angular scale of the scintillation pattern $\theta_{\rm cr} =r_{\rm diff}/2z$.  Equation (\ref{ACFthin}) was first derived for gaussian turbulence but it has also been derived approximately for Kolmogorov turbulence (\cite{Gwinn}).
The spectral autocorrelation in the thin screen model is degenerate to the combination $S$, $\Delta \nu_{\rm t}$ and $\theta_0/\theta_{\rm cr}$, so it is only possible to fit for two of these three parameters.  The uncertainty in the base level of the observed spectral autocorrelation function necessitates the introduction of the additional constant $A_{\rm off}$.  
When the source size exceeds the critical angular scale $\theta_{\rm cr}$ the form of the spectral decorrelation simplifies to
\begin{eqnarray}
F_{\rm thin} \approx A_{\rm off} + S_{\rm diff}^2 \left\{ \begin{array}{ll}
\left( \frac{\theta_0}{\theta_{\rm cr}} \right)^2, & \Delta \nu \la \Delta \nu_{\rm t} \frac{\theta_0}{\theta_{\rm cr}} \\
\left( 1 + \frac{\Delta \nu^2}{\Delta \nu_{\rm t}^2} \right)^{-1}, & \Delta \nu \ga \Delta \nu_{\rm t} \frac{\theta_0}{\theta_{\rm cr}} \\
\end{array} \right. . 
\end{eqnarray}
The characteristic scale of the frequency pattern is thus set by the source size to 
\begin{eqnarray}
\Delta \nu = \Delta \nu_{\rm t} \left( \frac{\theta_0}{\theta_{\rm cr}} \right). \label{nuratio}
\end{eqnarray}
The scale of the observed diffractive scintillation pattern is 
\begin{eqnarray}
s_0  = r_{\rm diff}  \sqrt{1+ \left( \frac{\theta_0}{\theta_{\rm cr}} \right)^2 }.  \label{S0thin}
\end{eqnarray}
This scale can be equated directly to the product of the diffractive scintillation timescale and velocity, $\viss \, t_{\rm diff}$.  A notable feature of the thin-screen model is that the pattern scale grows arbitrarily large with source size.

The spectral decorrelation associated with the extended medium model is (\cite{Chashei}), 
\begin{eqnarray}
F_{\rm ex} &=& A_{\rm off} + S_{\rm diff}^2 \, R(\Delta \nu) \left[1 + f(\Delta \nu) \frac{\theta_0^2}{\theta_{\rm cr}^2} \right]^{-1},  \label{ACFthick}
 \\
\hbox{where } &\null& f(\Delta \nu) =  2 \left(\frac{\Delta \nu_{\rm ex}}{\Delta \nu} \right)^{3 \over 2} 
   \frac{\sinh \left( \sqrt{\frac{\Delta \nu}{\Delta \nu_{\rm ex}} } \right) -
    \sin \left( \sqrt{\frac{\Delta \nu}{\Delta \nu_{\rm ex}} } \right)  }{ \cosh \left( \sqrt{\frac{\Delta   
    \nu}{\Delta \nu_{\rm ex}} } \right) +
    \cos \left( \sqrt{\frac{\Delta \nu}{\Delta \nu_{\rm ex}} } \right)  }, 
\nonumber \\ &\null& R(\Delta \nu) = 2 \left[  \cosh \left( \sqrt{\frac{\Delta \nu}{\Delta \nu_{\rm ex}} } \right) + 
\cos \left( \sqrt{\frac{\Delta \nu}{\Delta \nu_{\rm ex}} } \right) \right]^{-1},
\end{eqnarray}
where the point-source decorrelation bandwidth is $\Delta \nu_{\rm ext}= \pi \nu k r_{\rm diff}^2/\Delta z$.  When the source is extended the spectral decorrelation function is nearly degenerate to a combination of the free parameters, but it takes the following simple form
\begin{eqnarray}
F_{\rm ex} \!\!\! &\approx& \!\!\!  A_{\rm off} \!+\! S_{\rm diff}^2 \!\! \left\{ \begin{array}{ll}
3 \left( \frac{\theta_{\rm cr}}{\theta_0^2} \right)^2, & \Delta \nu \ll \Delta \nu_{\rm ex} \\
\frac{R(\Delta \nu)}{2} \left( \frac{\Delta \nu_{\rm ex}}{\Delta \nu} \right)^{3 \over 2} \left( \frac{\theta_{cr} }{\theta_0} \right)^2 , & \!\!\Delta \nu_{\rm ex} \la \Delta \nu \la \Delta \nu_{\rm ex} \left( \frac{\theta_0}{\theta_{cr}} \right)^{4 \over 3}  \\
R(\Delta \nu), & \Delta \nu \gg \Delta \nu_{\rm ex} \left( \frac{\theta_0}{\theta_{cr}} \right)^{4 \over 3}. \\
\end{array} \right. 
\end{eqnarray}
Source size reduces the overall amplitude of the spectral autocorrelation function by a factor $\theta_0^2/\theta_{\rm cr}^2$.
The diffractive pattern scale is largely insensitive to source size, and grows to a maximum of only twice the diffractive scale length as the source size increases:
\begin{eqnarray}
s_0 = r_{\rm diff} \sqrt{ \frac{1+\frac{1}{3} \left( \frac{\theta_0}{\theta_{\rm cr}}\right)^2 }{1+\frac{1}{12} \left( \frac{\theta_0}{\theta_{\rm cr}} \right)^2 } }. \label{S0thick}
\end{eqnarray}

Both models assume that $\theta_0/\theta_{\rm cr}$ is a constant and do not account for any possible variation of source size relative to the critical angle, $\theta_{\rm cr}$, with frequency.   Both also assume a circularly symmetric source and an isotropic scattering medium, the implications of which are discussed further below.

\subsubsection{Simple Brightness Temperature Estimate} \label{SimpleEst}
\begin{figure*}[h] 
\centerline{\psfig{file=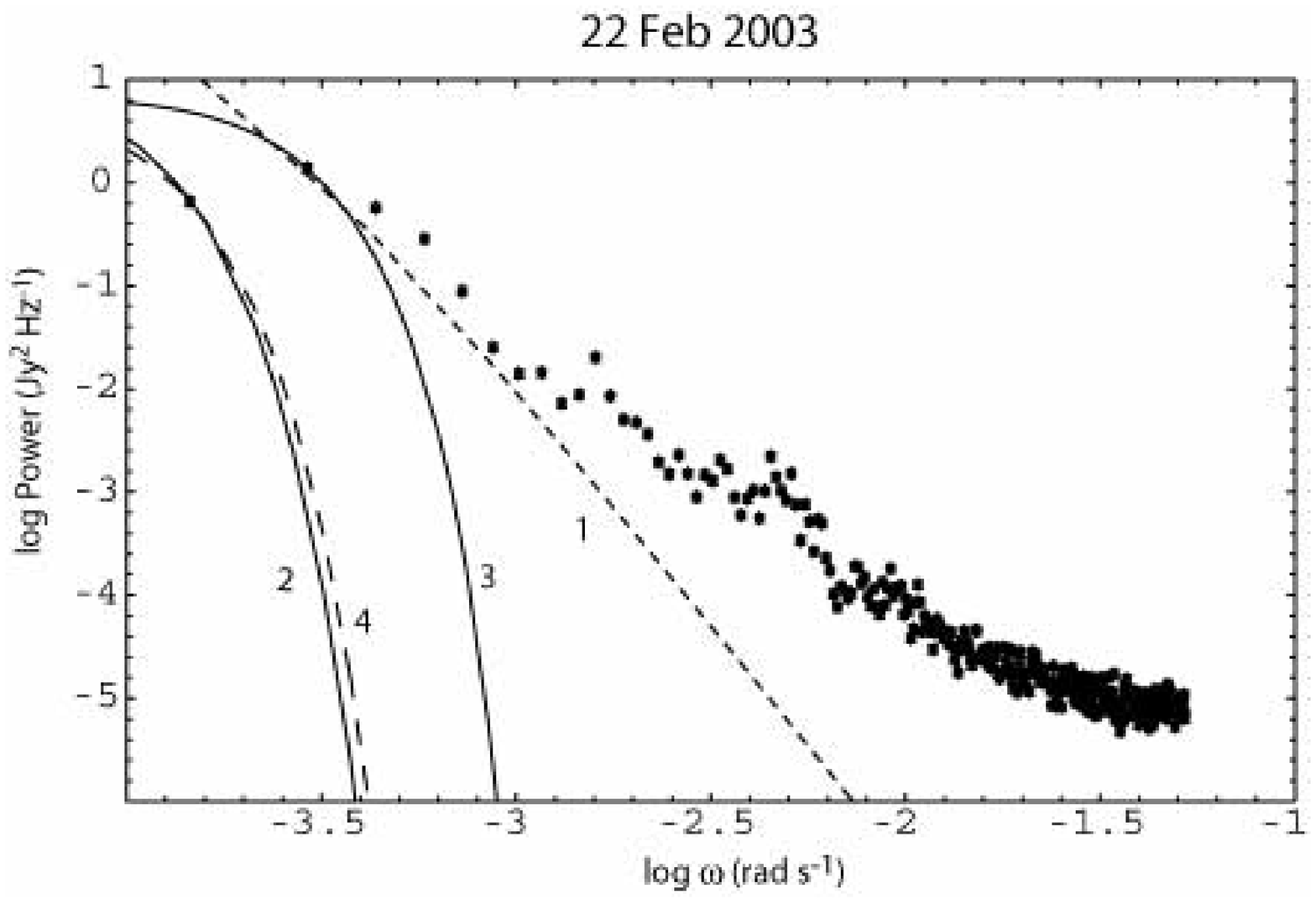,width=110mm}}
\centerline{\psfig{file=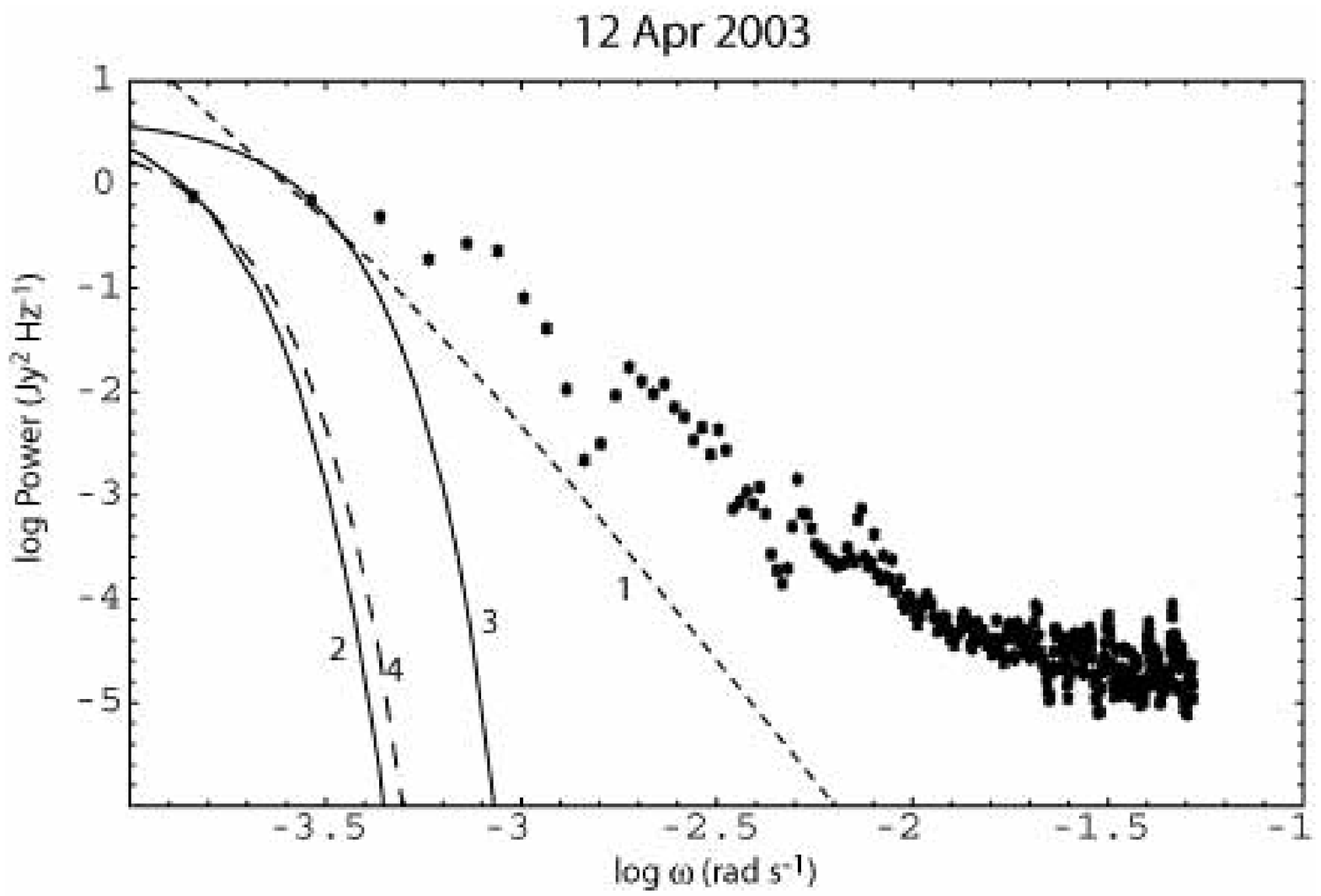,width=110mm}}
\caption{The average power spectra of the temporal variations on 22
Feb 2003 (top) and 12 Apr 2003 (bottom) shown with the expected thin
screen and extended medium power spectra for comparison.  The lines
are labelled according to the model: (1) best-fitting extended medium
with medium scale length 15\,pc, (2) thin screen model with $z=15\,$pc
with $C_N^2 \Delta L$ adjusted to match the power at the lowest
frequency bin (3) same as (2) but with $C_N^2 \Delta L$ adjusted to
match the power at the second-lowest frequency bin and (4) a thin
screen model with $z=4\,$pc with $C_N^2 \Delta L$ chosen to match the
power at the lowest frequency bin.  A scintillation speed
$\viss=50\,$km\,s$^{-1}$ is assumed.  Spectra matched to the second
lowest point are shown because they exhibit more power at high
frequencies.  However, even for this conservative choice they do not
replicate the observed power at high frequencies. In all models all of
the source flux density is assumed to undergo scintillation.  For the
thin-screen model a point source is assumed; a source of finite size
further decreases the power at high frequency.  In the extended medium
model used here, the form of the power spectrum is valid for
$\theta_{\rm src} > \theta_{\rm scat}$, so we choose $\theta_{\rm src}
=1.0\,$mas.  However, the curves shown here differ little from ones in
which the model is stretched by using $\theta_{\rm src} =0.1\,$mas
instead.  The parameters of the fits are summarised in Table 1.  A
minimum bias filter (Papoulis 1991, pp445-455) has been applied to the
spectra in order to reduce the errors inherent in the spectral
estimation, however the error bars shown are not mutually independent.}
\label{PowerSpectraFig}
\end{figure*}

The size of the feature associated with the narrowband intensity
fluctuations can be estimated in a simple manner for the thin screen
model given the transition frequency and the distance to the
scattering screen.  The transition frequency, $\nu_t$, can be deduced
from fits to the refractive power spectrum given the distance to the
scattering medium and the scintillation speed (Fig.\,\ref{PowerSpectraFig} and Table 1).
Dennett-Thorpe \& de Bruyn (2003) argue that the screen distance is in
the range $z=4-12\,$pc.  We estimate the brightness temperature for
screen distances of 4 and 15\,pc assuming a scintillation
speed of $\viss=50\,$km\,s$^{-1}$, a value deduced on the basis of
measurements at 6\,cm (Dennett-Thorpe \& de Bruyn 2003).

The amplitude of the frequency-dependent scintillation is $20\,$mJy
and its decorrelation bandwidth is $170$\,MHz (see Fig.\,\ref{ACFFigs}).  The flux
density associated with the scintillating component is $S_\nu \approx
20/(\theta_0/\theta_{\rm cr})$\,mJy.  The ratio $(\theta_0/\theta_{\rm cr})$ is estimated 
directly from the ratio of the
observed to point source decorrelation bandwidth by employing
eq. (\ref{nuratio}).  This equation is valid in the present case since
the observed decorrelation bandwidth is several times larger than the
point source decorrelation bandwidth for the expected value of $\nu_t
\ga 4$\,GHz (see Table 1).  The critical angular scale $\theta_{\rm
cr}$ is estimated directly from the scattering screen distance and
the transition frequency.  One solves for $\theta_0$ using
$\theta_{\rm cr}$ and the ratio $\theta_0/\theta_{\rm cr}$ to
obtain,

\begin{eqnarray}
\theta_{\rm src} = 7.4 \,\left( \frac{z}{1\,{\rm pc}} \right)^{-1/2}
\left( \frac{\nu_t}{1\,{\rm GHz}} \right)^{17/10} \,\mu{\rm as}.
\end{eqnarray}

We use the fits to the refractive power spectrum in Fig.\,\ref{PowerSpectraFig} to estimate
the transition frequency.  For $z=15\,$pc one has $\nu_t=3.89\,$GHz
and $r_{\rm diff} = 2.19 \times 10^7$\,m at 1.4\,GHz and the expected
decorrelation bandwidth of a point source at 1.4\,GHz is 43.3\,MHz.
The corresponding numbers for a screen at $z=4\,$pc are
$\nu_t=5.45\,$GHz, $r_{\rm diff} = 6.39 \times 10^6\,$m and $\nu_{\rm
dc}=13.8$\,MHz.  For $z=15\,$pc the source size is $\theta_0 =
19\,\mu$as, while for $z=4\,$pc the source size is $67\,\mu$as.

The transition frequency cancels out of the expression for the source
brightness temperature leaving,
\begin{eqnarray}
T_B = \frac{\lambda^2 S_\nu}{2 k_B \pi \theta_0^2 } (1+z_S) =
4.9 \times 10^{12} \left( \frac{z}{1\,{\rm pc}} \right) \,{\rm K},
\end{eqnarray}
where we correct the brightness temperature for the source redshift
$z_S=0.54$ and we take $\lambda =0.21\,$m.  For $z=15\,$pc the implied
brightness temperature is $7 \times 10^{13}\,$K while for $z=4\,$pc it
is $2 \times 10^{13}\,$K.

Two factors account for the higher brightness temperature estimated at
1.4\,GHz relative to that at 4.9\,GHz: (i) the wavelength is 3.5 times
larger and (ii) for screen distances $z \approx 15\,$pc, the source
size is estimated to be up to $\sim 3$ times smaller than the value
estimated at 6\,cm.


\subsubsection{Fits to the spectral autocovariance}  \label{detailed}
For completeness we also estimate the brightness temperature using
parameters extracted from the fits to the spectral autocovariance,
without imposing any external constraints (e.g. from the fits to the
refractive power spectrum).

Fits to the frequency autocovariance functions are shown in Fig.\,\ref{ACFFigs}.
Both models provide close fits to the data, with reduced
$\chi^2$ values less than unity.  Fit parameters and confidence limits
for both thin-screen and extended medium models are listed in Table 2.
We note that strong RFI causes difficulties in the spectral
normalisation of the 12 Apr 2004 dataset, which introduces systematic
errors in the frequency autocovariance function and the scintillation
parameters derived from it.  The scintillation and source parameters
derived from this dataset should be treated with caution.

In the thin screen model we fit for the products $\Delta \nu_{\rm t}
S_{\rm diff}$ and $S_{\rm diff}^{-1} \theta_0/\theta_{\rm cr}$,
leaving the flux density of the scintillating component as a free
parameter.  In the extended medium model we fit to the combination
$S_{\rm diff}^2 \theta_{\rm cr}^2/\theta_0^2$ because there is a strong
degeneracy between between $S_{\rm diff}$ and $\theta_0/\theta_{\rm
cr}$ once the source size exceeds $\theta_{\rm cr}$.

Equations (\ref{ACFthin}) and (\ref{S0thin}) or (\ref{ACFthick}) and
(\ref{S0thick}) are used in conjunction with the fit parameters and
scintillation velocity and timescale to derive the screen distance,
source size and brightness temperature for either thin- or
thick-screen models.  These derived source parameters are listed in
Table 3.

\begin{figure*}[h] 
\begin{center}
\centerline{\psfig{file=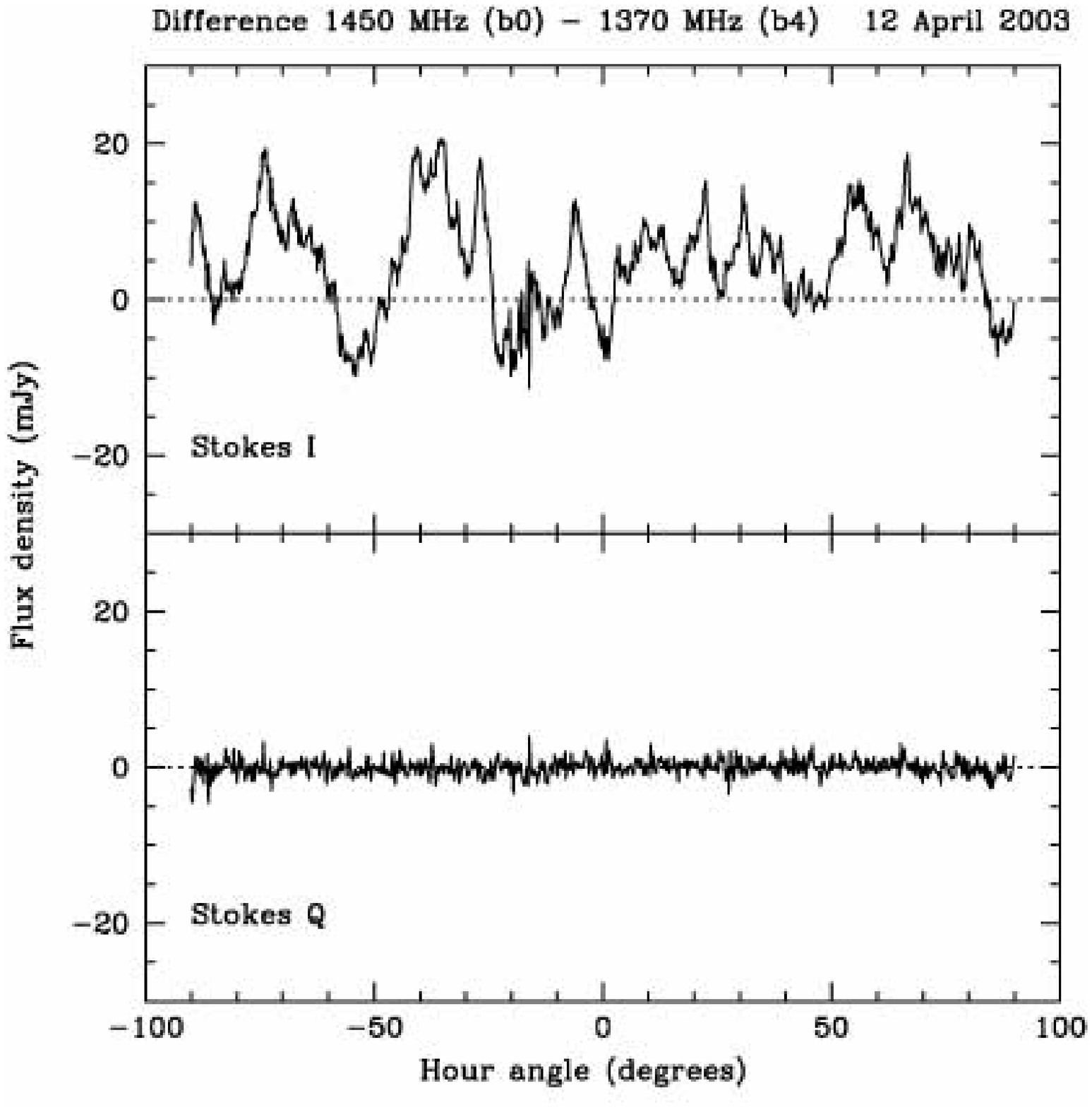,width=70mm}  \psfig{file=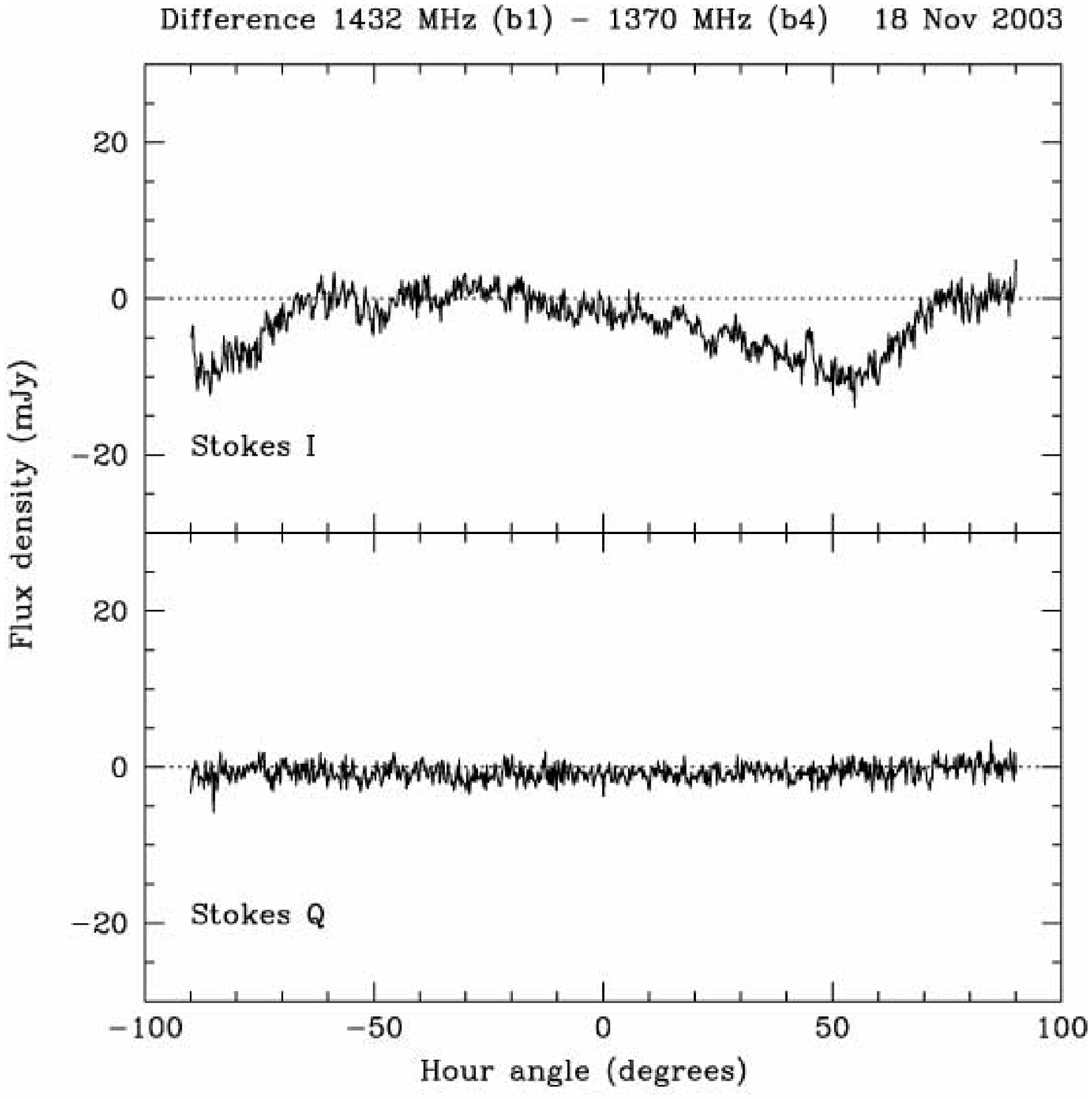,width=70mm} }
\begin{tabular}{c}
(b) \psfig{file=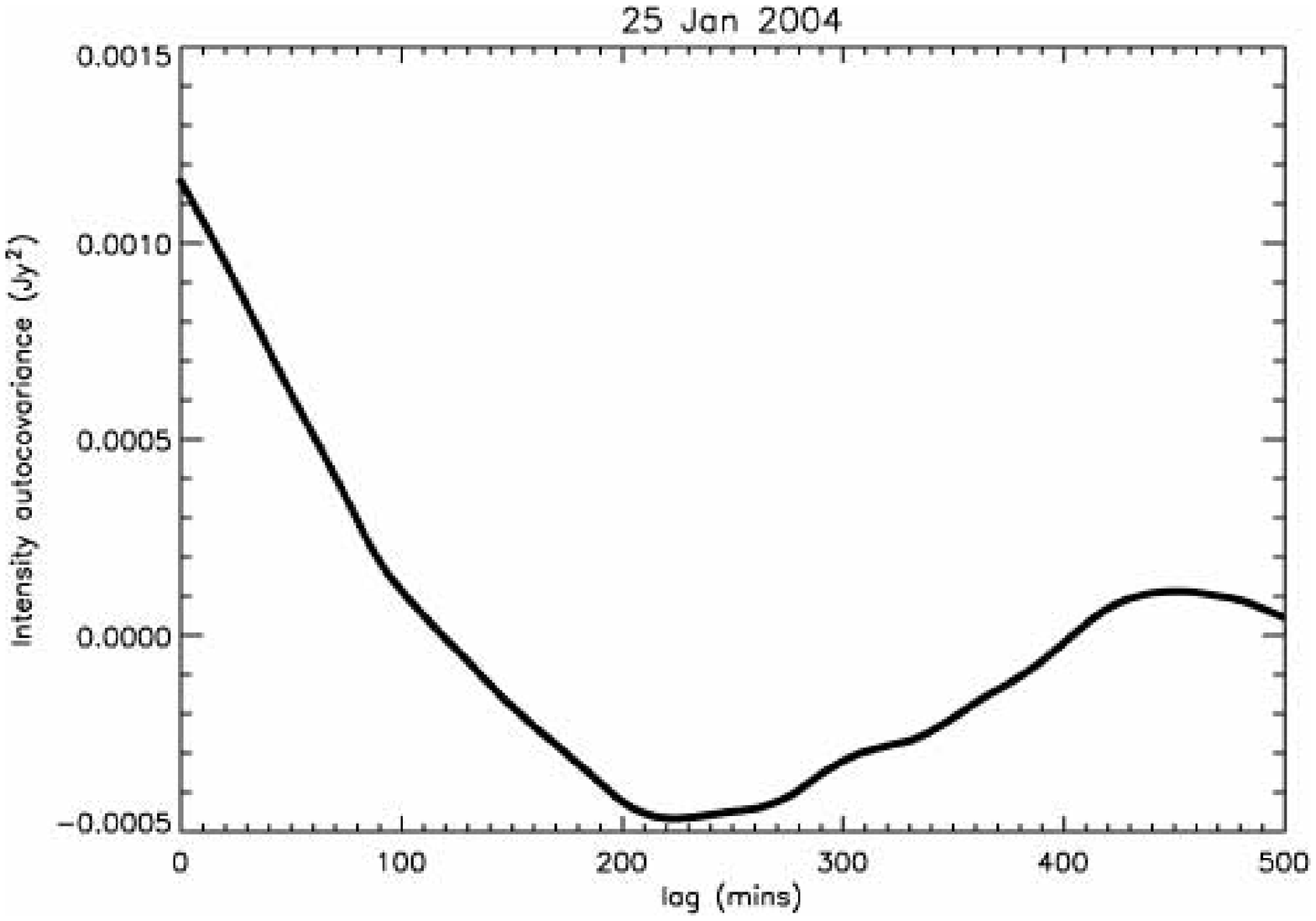,width=60mm} \\
\psfig{file=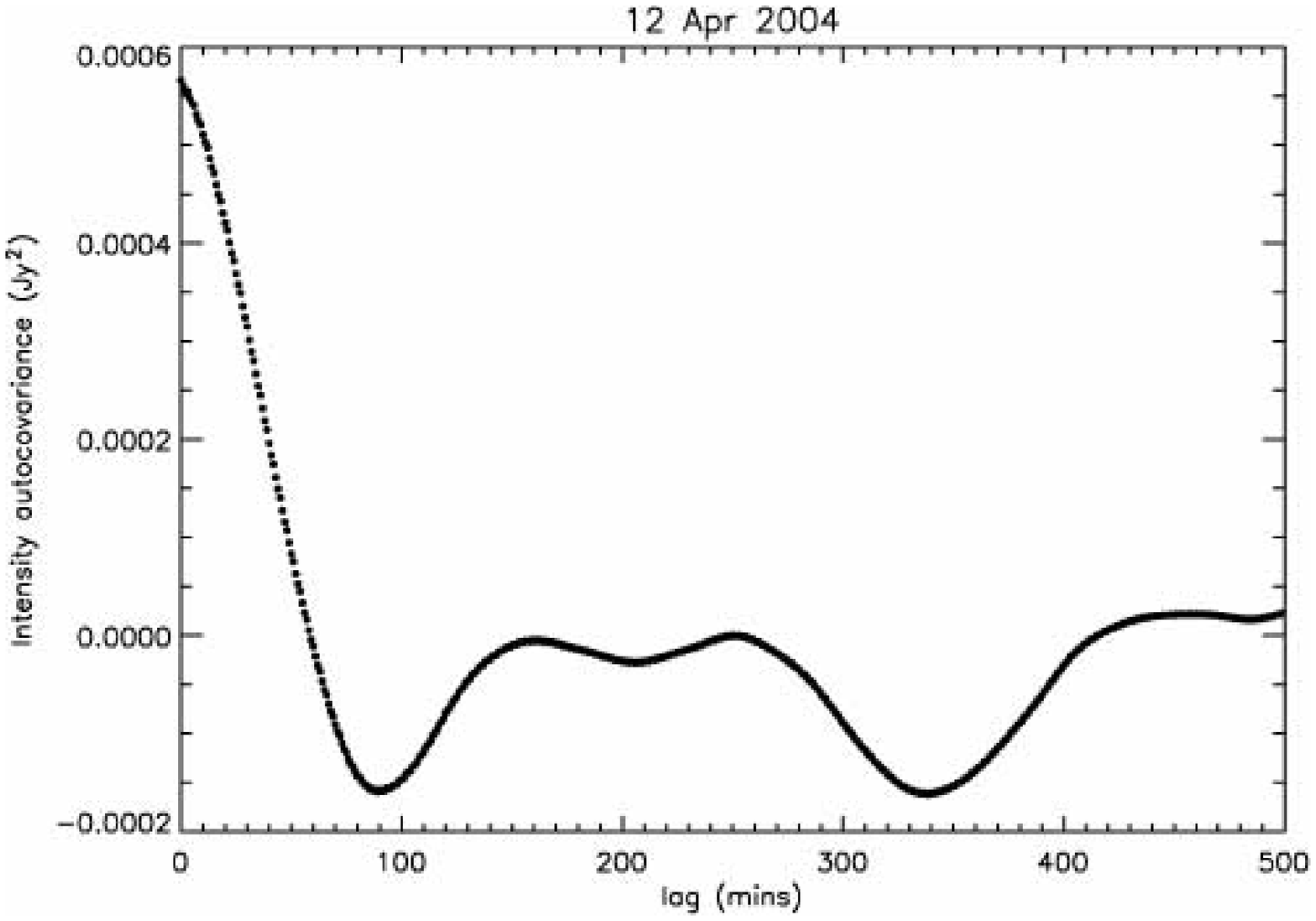,width=60mm} \\
\psfig{file=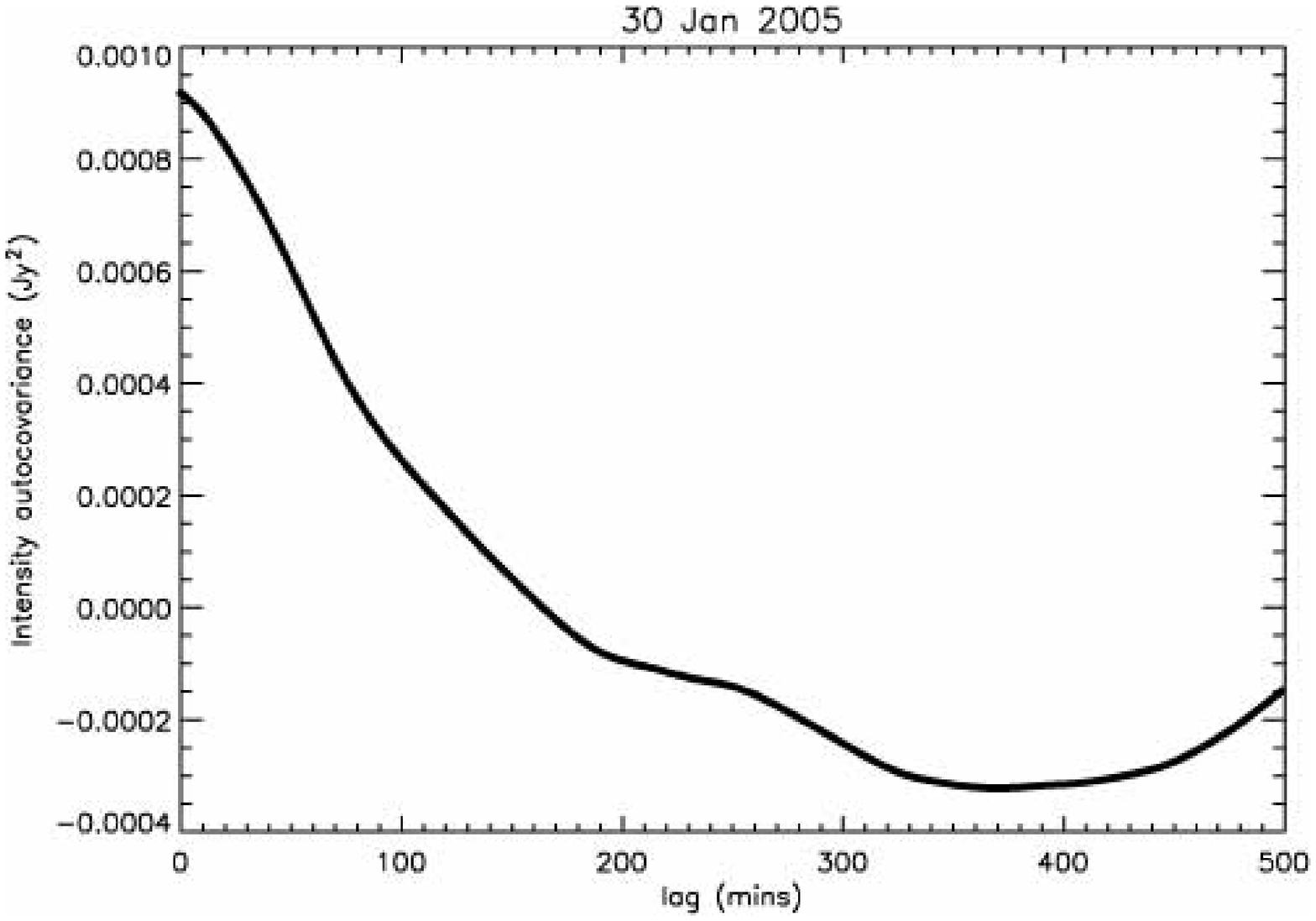,width=60mm} \\
\end{tabular}
\end{center}
\caption{(a) Two sample difference light curves, from 12 Apr 2003 (fast
season) and 18 Nov 2003 (slow season).  The rapid, sharp variations of
these frequency-dependent intensity fluctuations, particularly on 12
Apr 2003, are well above the 1.4\,mJy noise level.
Note that the rapid frequency-dependent fluctuations are absent in the 
slow season.  
(b) Autocovariance functions of the temporal intensity fluctuations on
25 Jan 2004, 12 Apr 2004, 30 Jan 2005.  Note that the amplitude of the
function at zero temporal lag significantly exceeds the amplitude
observed in the spectral autocovariance function (see
Fig.\,\ref{ACFFigs}).  The excess can be attributed to the additional
influence of refractive scintillation on the temporal modulations.}
\label{DiffLightFigs}
\end{figure*}

Many of the quantities derived in Table 3 depend on the scintillation
timescale.  The scintillation timescale at each epoch is determined
by computing the intensity autocovariance function for each 17.5\,MHz
band and measuring the point at which this falls to $1/e$ of its
maximum value.  The mean timescale at each epoch is computed by
averaging the timescales derived for the various bands.  The
scintillation timescales are $67\pm 3$\,min, $40\pm 2$\,min and
$87\pm 3$\, for 22 Feb 2004, 12 Apr 2004 and 30 Jan 2005 respectively.
The quoted error in the timescale is the standard error of the mean,
computed from the variation in timescale observed between bands.  The
band-averaged temporal autocovariance functions for these dates are
shown in Fig.\,\ref{DiffLightFigs}.  In Table 3 the scintillation timescale is expressed
as a multiple of 1\,hour.

We reject the extended medium model because its estimates of the
scattering medium properties are unphysical.  The large medium depth
indicated by this model would place most of the scattering medium well
outside the Galactic plane.  Formally, the estimated depth is large
because, for an extended source scintillating in an extended medium,
the diffractive pattern scale $s_0$ asymptotes to twice the
diffractive scale length $r_{\rm diff}$.  This fixes the diffractive
scale length to a much larger value relative to the thin screen model,
in which the pattern scale increases with source size without bound.
Thus, in the extended medium model, one requires a large medium depth
for a given decorrelation bandwidth since the latter is proportional to $r_{\rm diff}^2 /z$.  It
should be noted that just such a misestimate is expected to occur
when the source is extended but is in fact subject to scattering
through a thin screen.

We regard the range of the brightness temperatures derived at
different epochs as the most faithful estimate of their true
uncertainty.  Part of the range can be attributed to the uncertainty
in the scattering speed and scintillation timescale, both of which
change between epochs.  The minimum brightness temperature is
uncertain by a factor of two even when reasonable variations in these
parameters are taken into account and the RFI-afflicted 12 Apr 2004
dataset is excluded.  The uncertainty may reflect other uncertainties
not taken into account by the model, such as anisotropy in the source
and scattering medium.

We note that the nearby pulsar PSR\,1813$+$4013 is observed to 
exhibit a diffractive
decorrelation bandwidth of $\sim \! 10\,$MHz at 1.4\,GHz (B. Stappers, private communication).  If
the scattering properties of J1819$+$3845 are comparable then this
decorrelation bandwidth favours a value of $S_{\rm diff}$ around 100\,mJy, and a
brightness temperature toward the low end of its allowed range.

\begin{table*}
\begin{center}
\begin{tabular}{| c  |c | c | c | c |}
\hline
\null & fit parameter & 25 Jan 2004  & 12 Apr 2004$^\dagger$ & 30 Jan 2005 \\ \hline
\null & offset $A_{\rm off}$ (Jy$^2$) & $-3.46 \pm 0.03 \times 10^{-4}$   & $-2.98 \pm 0.04 \times 10^{-4}$ 
& $-2.93 \pm 0.02 \times 10^{-4}$ \\
thin screen & bandwidth $\Delta \nu_{\rm t} S_{\rm diff}$ (MHz\,Jy) & $4.09 \pm 0.05$  & $4.68 \pm 0.07$ & $3.89 \pm 0.03$ \\  
\null & size $(1+\theta_0^2/ \theta_{\rm cr}^2) S_{\rm diff}^{-2}$ (Jy$^{-2}$) & $1740 \pm 10 $ & $1680 \pm 20$ & $1600 \pm 10$ \\ \hline
extended & offset $A_{\rm off}$ (Jy$^2$) & $-3.10 \pm 0.03 \times 10^{-4}$   & $-2.60 \pm 0.04 \times 10^{-4}$   & $-3.06 \pm 0.02 \times 10^{-4}$ \\
medium & component flux density and source size $S_{\rm diff}^2 \theta_{\rm cr}^2/\theta_0^2$ (Jy$^2$) & $1.78 \pm 0.01 \times 10^{-4}$ &  $1.81 \pm 0.02 \times 10^{-4}$ 
& $2.11 \pm 0.01 \times 10^{-4}$  \\
\null & bandwidth $\Delta \nu_{\rm ext}$ (MHz) & $5.91 \pm 0.07 $ & $ 6.7 \pm 0.1$ & $5.91 \pm 0.06$ \\ \hline
%
\end{tabular}
\end{center}
\caption{Fit parameters with formal 1$\sigma$ errors from the fit. 
$\dagger$Strong RFI affected the spectral normalization of the 12 Apr 2004 observation, which in turn affected estimation of the frequency autocovariance upon which this fit is based.  }
\end{table*}

\begin{table*}
\begin{center}
\begin{tabular}{|c|c|c|c|c|c|}
\hline
model & quantity & scaling & 25 Jan 2004 & 12 Apr 2004 & 30 Jan 2005 \\ \hline
\null & screen distance (pc) & $S_{\rm diff}^{-1} t_{1hr}^2 {\rm v}_{50}^2$ & $ 6.0 $ & $ 5.5 $ & $ 6.9 $ \\
\null &   source size ($\mu$as) ($S_{\rm diff}=0.05\,$Jy) & $t_{1hr}^{-1} {\rm v}_{50}^{-1}$ & $ 4.4 $ & $ 4.8 $ & $ 3.8 $ \\
thin screen &   source size ($\mu$as) ($S_{\rm diff}=0.15\,$Jy) & $t_{1hr}^{-1} {\rm v}_{50}^{-1}$ &  $ 15 $ & $ 16 $ & $ 13 $ \\
\null & brightness temperature (K)($S_{\rm diff}=0.05\,$Jy) &  $t_{1hr}^2 {\rm v}_{50}^2$  & $9.1  \times 10^{14} $ & $7.6 \times 10^{14}$ & $12 \times 10^{14} $ \\ 
\null & brightness temperature (K)($S_{\rm diff}=0.15\,$Jy) & $ t_{1hr}^2 {\rm v}_{50}^2$ & $2.4  \times 10^{14}$ & $1.9 \times 10^{14} $ & $3.1 \times 10^{14}$ \\ \hline
\null & medium thickness (kpc) & $t_{1hr}^2 {\rm v}_{50}^2$ & $ 5.7 $ & $7.2 $ & $5.7 $  \\
extended medium & source size ($\mu$as) &  $S_{\rm diff}\, t_{1hr}^{-1} {\rm v}_{50}^{-1}$ & $ 3.9 $ & 
$ 3.1  $ & $ 3.6$   \\
\null & brightness temperature (K) & $S_{\rm diff}^{-1} \, t_{1hr}^2 {\rm v}_{50}^2$ & $ 1.5 \times 10^{16}  $ & $ 2.4 \times 10^{16} $ &  $ 1.7 \times 10^{16} $ \\ \hline
\end{tabular}
\end{center}
\caption{Source parameters derived from the best-fit parameters of the
various scintillation models applied to the 25 Jan 2004, 12 Apr 2004
and 30 Jan 2005 observations.  A range of parameters are permitted by
the thin screen model because the diffractively scintillating flux
density is unknown.  A larger flux density implies a greater source
size and lower brightness temperature.  The maximum possible flux
density is the intrinsic flux density of the entire source, which is
approximately 150\,mJy.  The numbers in the three last columns should
be multiplied by the scaling parameter to derive the correct value of
each quantity.  Here $S_{\rm diff}$ is measured in Jansky, $t_{1hr}$
is the diffractive scintillation timescale in hours and $v_{50}$ is
the scintillation speed normalised to 50\,km\,$s^{-1}$.}
\end{table*}

\subsubsection{Source spectral changes}

In the scintillation model above the brightness temperature depends on the free parameter $S_{\rm diff}$.
We discuss here the extent to which the spectrum of the component undergoing scintillation matches the mean source spectrum, and whether this component makes a significant contribution to the overall intrinsic source spectrum at low frequencies.  The latter might be expected if the source is comprised of multiple components with distinct spectra.  Evidence that this may be case comes from analysis of the 4.9 and 8.4\,GHz light curves, in which distinctly different polarization and total intensity fluctuations imply that the source is composed of at least two bright features (Macquart, de Bruyn \& Dennett-Thorpe 2003).


Despite the source's complex structure, its mean spectrum between 1.4
and 8.4\,GHz is a power law with a spectral index of 0.8.  We have
also measured the spectral index of the source from the variation in
mean flux density across the band from our 21\,cm observations.  These
measured spectral indices, listed in Table 4, varying between 0.8 and
1.2, are, with one significant exception, {\it consistent} with the
intrinsic spectral index derived on the basis of long-term
measurements between 1.4 and 8.4\,GHz (de Bruyn et al., in prep.).  It
is difficult to be more precise, since our 1.4\,GHz spectral
measurements are a poor indicator of the intrinsic source spectrum
when only a few diffractive scintles are observed, as is the case in
many of our observations.

The one notable exception is the spectrum measured on 22 Feb 2003,
which is wildly at variance with the mean source spectrum and with the
other observations.  This difference may be significant, because the
average spectrum extracted from this observation encompasses many
diffractive scintles.  This difference may reflect the emergence of a
new component in the source.  However, no discernible deviation in the
amplitude of diffractive scintillation is associated with this epoch.


\begin{table*}
\begin{center}
\begin{tabular}{| l |c|c|}
\hline
date & mean flux density (mJy) & spectral index $\alpha$ \\ \hline
14 Jun 2002 & 85 & $0.79 \pm 0.01$ \\
30 Aug 2002 & 80 & $-$ \\
22 Feb 2003 & 144 & $0.29 \pm 0.01$ \\
12 Apr 2003 & 111 & $1.01 \pm 0.01$ \\
19 Jun 2003 & 115 & $1.19 \pm 0.01$ \\
22 Aug 2003 & 242 & $(1.21 \pm 0.02)$ \\
18 Nov 2003 & 151 & $-$ \\
25 Dec 2003 & 147 & $1.05 \pm 0.01$ \\
22 Feb 2004 & 144 (at 1.40 GHz)& $ 0.90 \pm 0.06 $ \\
12 Apr 2004 &  143 (at 1.40 GHz)& $ 0.67 \pm 0.02$ \\
30 Jan 2005 & 167 (at 1.40 GHz)& $ 0.50 \pm 0.03 $ \\
\hline
\end{tabular}
\end{center}
\caption{The variation in mean flux density and spectral index of J1819$+$3845 with observing date. Here the spectral index $\alpha$, defined as $S\propto \nu^\alpha$, is derived solely on the basis of the spectrum exhibited across the band at 1.4\,GHz.  Blanks and bracketed values indicate observations in which so few diffractive scintles are present that the mean spectrum is a poor representation of the mean source spectrum, and blanks indicate instances for which a power law is an unacceptable fit to the mean spectrum.  The errors quoted in the spectral index reflect only formal errors associated with a fit to a power law.}
\end{table*}


\section{Discussion}

\subsection{Robustness of the brightness temperature estimate}

The $\ga10^{14}\,$K brightness temperature implied by the diffractive
scintillation properties of J1819$+$3845 is difficult to account for
using the standard interpretation of AGN radio emission in terms of
synchrotron emission.  In this section we consider the robustness of
this estimate.

The greatest source of error in the thin-screen model is associated
with the effect of anisotropy on the scale of the scintillation
pattern, $s_0$, which propagates into the estimation of $r_{\rm
diff}$.  Scattering measurements of pulsars (Mutel \& Lestrade 1990;
Spangler \& Cordes 1998) suggest that the maximum degree of anisotropy expected
due to turbulence in the interstellar medium is $3:1$.  However,
intensity variations in the regime of weak scattering at 4.9\,GHz
indicate that the scintillation pattern of J1819$+$3845 has an axial
ratio of $14_{-8}^{+>30}$ (Dennett-Thorpe \& de Bruyn 2003).  The
relative contributions of medium and source to this overall anisotropy
are unknown at this frequency.  At 1.4\,GHz the source structure is
expected to be the primary agent responsible for any anisotropy in the
scintillation pattern because the source substantially exceeds the
critical angular scale of the diffraction pattern.  We estimate $3.2
< \theta_0/\theta_{\rm cr} < 37 $ for $50 < S_{\rm diff} < 150\,$mJy
(see Table 2).

Anisotropy in the source must couple with anisotropy intrinsic to the
turbulence in the scattering medium to cause an appreciable
misestimate of the source size.  This is because any source
elongation oriented parallel to the medium anisotropy would not be
detected by a reduction in the scintillation amplitude.  In this case
an anisotropic ratio of $\zeta$ would lead to a misestimate of the
source size by a factor of $\zeta$ along one source axis and a
brightness temperature misestimate of a factor of $\zeta$.  Anisotropy
in the source alone is insufficient to cause a serious misestimate of
its brightness temperature, because the source extension would
manifest itself through a reduction in the modulation amplitude of the
scintillation; although the source is assumed to be circularly
symmetric in our fits above, the reduction of the modulation amplitude
would lead us to deduce a source angular size somewhere between the
lengths of the short and long axes of the source.

We have performed an analysis of the variation in scintillation time
scale at 21\,cm.  An annual cycle in the timescale is clearly
observed, but anisotropy in the scintillation pattern is not required
to reproduce the timescale variations observed from our observations
to date.  Changes in the magnitude of the scintillation velocity, as
the Earth's velocity changes with respect to the scattering medium's,
are alone sufficient to reproduce the annual cycle.


Another shortcoming of the scattering models resides in the assumption
that the scattering occurs in the regime of asymptotically strong
scintillation.  The scattering strength $r_{\rm F}/r_{\rm diff}$ is
derived from the decorrelation bandwidth.  The decorrelation bandwidth
in the thin-screen model is at most $\Delta \nu_t=78$\,MHz for $S_{\rm diff}=50\,$mJy,
which implies a scattering strength $\approx 4.2$.  For $S_{\rm
diff}=150\,$mJy the decorrelation bandwidth is $26$\,MHz, implying a
scattering strength of $7$ .  The scintillation is sufficiently strong
to be applicable to the present situation.  Certainly, any errors introduced by
this approximation are minor relative to those introduced by possible
anisotropy in the scintillation pattern.



\subsection{Relationship to 6\,cm source structure and scattering properties}

It is important to consider how the structure of the source derived
here relates to that inferred on the basis of the weak scintillation
exhibited by the source at 4.9\,GHz.

The source angular size derived at 1.4\,GHz for any given screen
distance (c.f. Sect. \ref{SimpleEst}) is approximately three times
smaller than that inferred at 4.9\,GHz.  This would seem surprising
because one would expect that the source, whose spectrum does not fall
off as fast as that of a uniform synchrotron self-absorbed 
source ($\alpha$=+2.5) should be substantially larger at 1.4\,GHz.  
A straightforward comparison between the scintillation properties 
at 1.4\,GHz and 4.9\,GHz is complicated by these opacity 
effects.  Significant  parts of the source that are visible at
4.9\,GHz, and contribute to the observed scintillations, 
may well be hidden at 1.4\,GHz. On the other hand, the diffractively
scintillating component need only comprise a small fraction of the
total source emission at 6\,cm because weak and diffractive scintillation are
sensitive to structure on widely different angular scales.  Weak
scintillation responds to all structure on angular scales $\la
\theta_{\rm F}$ whereas diffractive scintillation produces a strongly
identifiable response specifically to structure on much smaller
angular scales, $\sim \theta_{\rm cr}$.  Thus the small structure
responsible for the diffractive scintillation at 1.4\,GHz may well be
present at 4.9\,GHz but the signature of its presence could be masked by
the dramatic variations due to the rest of the source. It is also
possible that the source has multiple components at both 1.4\,GHz and
4.9\,GHz with very different spectral indices. For example, a coherent
emitter (see the next section) could well have a very steep spectrum
which would contribute a negligible fraction of the emission at
4.9\,GHz.

Recent observations suggest that this may indeed be the case.  We may have
detected evidence for the component responsible for diffractive
scintillation in the weak scattering of the source at 4.9\,GHz.
Observations from Dec 2003 to Apr 2004 at 4.9\,GHz indicate the
emergence of 5-10\,mJy variations on a timescale of $< 15\,$min
superposed on the $\sim 200\,$mJy peak-to-peak, $\approx 40\,$min
variations normally observed at 4.9\,GHz at this time of year.  It is
possible that the scintillation at 1.4\,GHz, which is more sensitive
to fine structure, first detected the same feature which was
subsequently detected at higher frequencies.  We continue to monitor
the source and will return to this apparent evolution in the future.

The screen distance indicated by the model in Sect.\ref{detailed} is
larger than the $4-12$\,pc value estimated by Dennett-Thorpe \& de
Bruyn (2003) assuming isotropic turbulence to model the intensity
fluctuations observed in the regime of weak scattering at 4.9\,GHz.
The minimum distance implied by the present thin-screen model is $\sim
40 \, t_{1hr}^2 v_{50}^2\,$pc if $S_{\rm diff}=150\,$mJy.  An obvious
reason for this discrepancy is that anisotropy is not taken into
account in the estimate of the screen distance at 4.9\,GHz, and we
have no detection of anisotropy at 1.4\,GHz.

\subsection{Problems with high brightness temperature emission}

The high brightness temperature exhibited by a component of
J1819$+$3845 raises concerns regarding the interpretation of AGN
emission in terms of incoherent synchrotron radiation. Inverse Compton
scattering limits the brightness temperature of incoherent synchrotron
emission to $10^{12}\,$K (\cite{Kellerman}) but equipartition
arguments (\cite{Readhead}) suggest that the actual limit should be an
order of magnitude below this.  Bulk motion with a Doppler boosting
factor $\delta \ga 100$ is required to reconcile the observed
brightness temperature with its maximum possible rest-frame value.
Such high bulk motions are problematic because they imply unacceptably
high jet kinetic energies.  Synchrotron emission is also extremely
radiatively inefficient in this regime, and it is questionable whether
$\Gamma \ga 100$ motions are compatible with the hypothesis of
incoherent synchrotron radiation (\cite{Begelman}).

In view of the difficulties confronted by an explanation involving
synchrotron radiation, it is appropriate to consider whether a
coherent emission mechanism provides a more acceptable explanation of
the high brightness temperature.  The nature of coherent emission
requires the source to be composed of a large number of independently
radiating coherent `bunches', with individual brightness temperatures
far in excess of the $T_b \ga 10^{14}\,$K value derived here
(e.g. \cite{Melrose91}).  This is because the coherence volume of any
one coherent bunch is microscopic compared to the light-week
dimensions of the source.  Further, the short lifetime associated with
any individual coherent bunch would require emission from a large
number of independent subsources to explain the constancy of the
emission observed on 12-hour to 6-monthly timescales.

Coherent emission from each bunch is expected to be highly polarized.
The upper limit of 1\% overall source polarization at 1.4\,GHz limits
the polarization associated with the diffractive component from
$\approx 1$\%, for $S_{\rm diff}=150\,$mJy, to 3\%, for $S_{\rm
diff}=50\,$mJy.  This suggests that either the emission is efficiently
depolarized as it escapes the source or that the emission is
intrinsically unpolarized.  The latter would occur if the magnetic
field is highly disordered within the emission region, so that the
polarizations of individual coherent patches would be diluted when
averaged over the entire region.

Another important obstacle relates to the escape of extremely bright
emission from the source region.  Induced Compton scattering places an
extremely stringent limit on the thermal electron density of the
source: for a path length $L$ the electron density must satisfy
\begin{eqnarray}
n_e \ll \frac{1}{\sigma_T L} \left( \frac{T_b}{5 \times 10^9 \, K}
\right)^{-1}=2.4 \left( \frac{L}{1\,{\rm pc}} \right)^{-1} \left(
\frac{T_b}{10^{15}\,{\rm K}} \right)^{-1} \, {\rm cm}^{-3},
\end{eqnarray} 
for induced Compton scattering to be unimportant.  It is argued that
this density is incompatible with the high densities required to
efficiently generate coherent emission in the first place
(e.g. \cite{Coppi}).  This difficulty may be overcome by appealing to
a highly anisotropic photon distribution.  However, this explanation
is also problematic because such highly beamed emission acts like a
particle beam in exciting Langmuir waves which also scatter the
radiation (\cite{Gedalin}; \cite{Luo}).  This effect is the accepted
mechanism for the occultation of several eclipsing radio pulsars whose
radiation propagates through a relatively low density stellar wind.
Applied in the context of AGN, this effect would require unreasonably
low electron densities in the emission and ambient media to permit the
escape of coherent emission from the source (\cite{Levinson}).

Begelman, Ergun \& Rees (2005) have recently proposed an
electron-cyclotron maser model for the high brightness temperature
emission inferred in some IDV sources.  They discuss in much more
detail the difficulties associated with the escape of bright radiation
from a source.  They conclude that it is possible for the high brightness
radiation observed in J1819$+$3845 to escape, subject to
certain constraints on the location of the emission region.

\section{Conclusions}
We have detected the diffractive interstellar scintillation from the quasar J1819$+$3845 at 1.4\,GHz.   This detection is notable because it constitutes the first detection of this phenomenon in an AGN, and it implies that a component of the source must be extremely compact.   

These scintillations are analysed in the context of thin-screen and extended-medium models for the distribution of interstellar scattering material.  The timescale, bandwidth and amplitude of the variations at 21\,cm imply a brightness temperature  $\ga 10^{14}\,$K.  
\begin{acknowledgements}
The WSRT is operated by the Netherlands Foundation for Research in Astronomy (NFRA/ASTRON) with financial support by the Netherlands Organization for Scientific Research (NWO).  We thank Ben Stappers, Barney Rickett and Bill Coles for discussions and useful comments. 
\end{acknowledgements}


\begin{figure*}[h] 
\centerline{\psfig{file=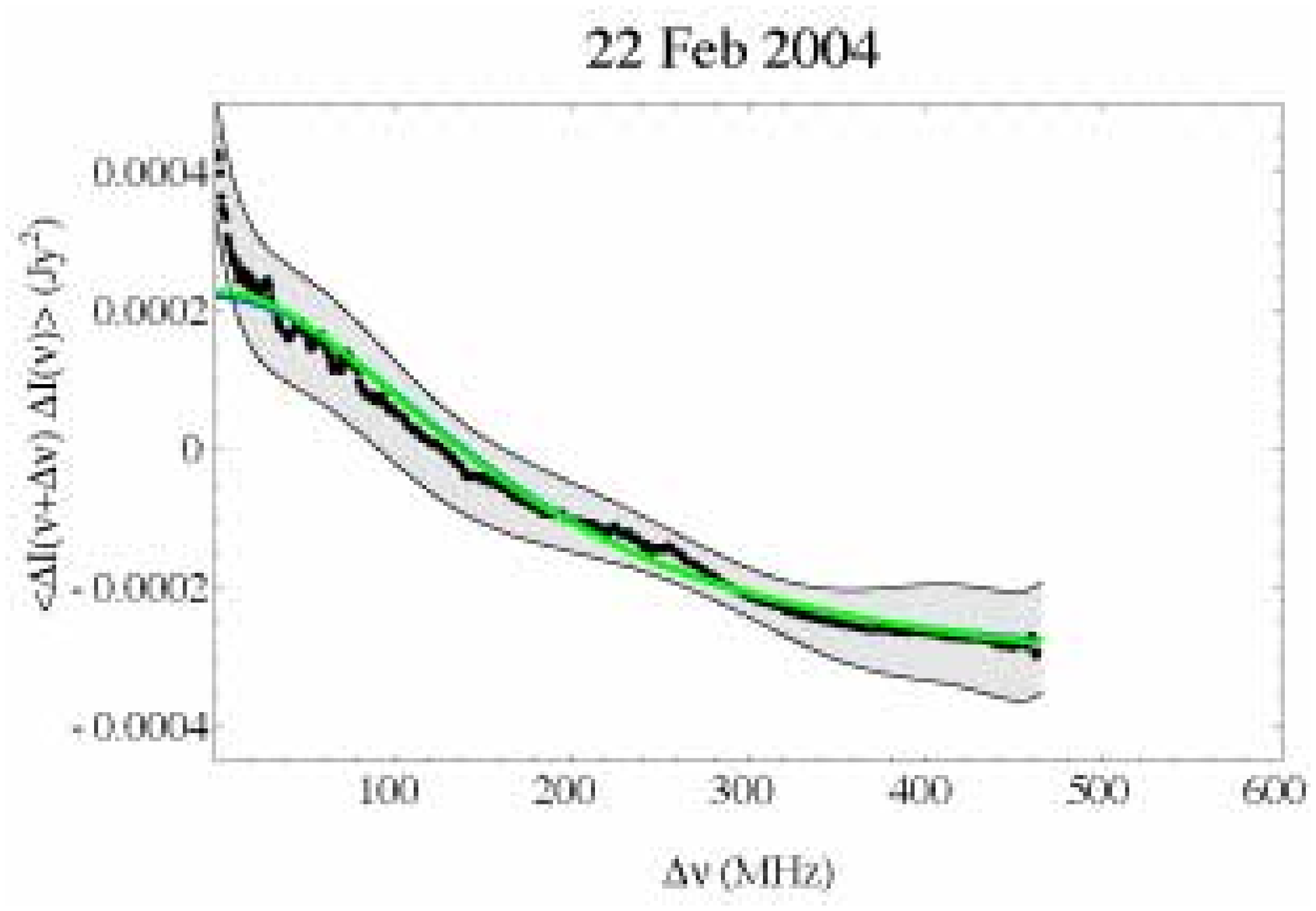,width=110mm}} 
\centerline{\psfig{file=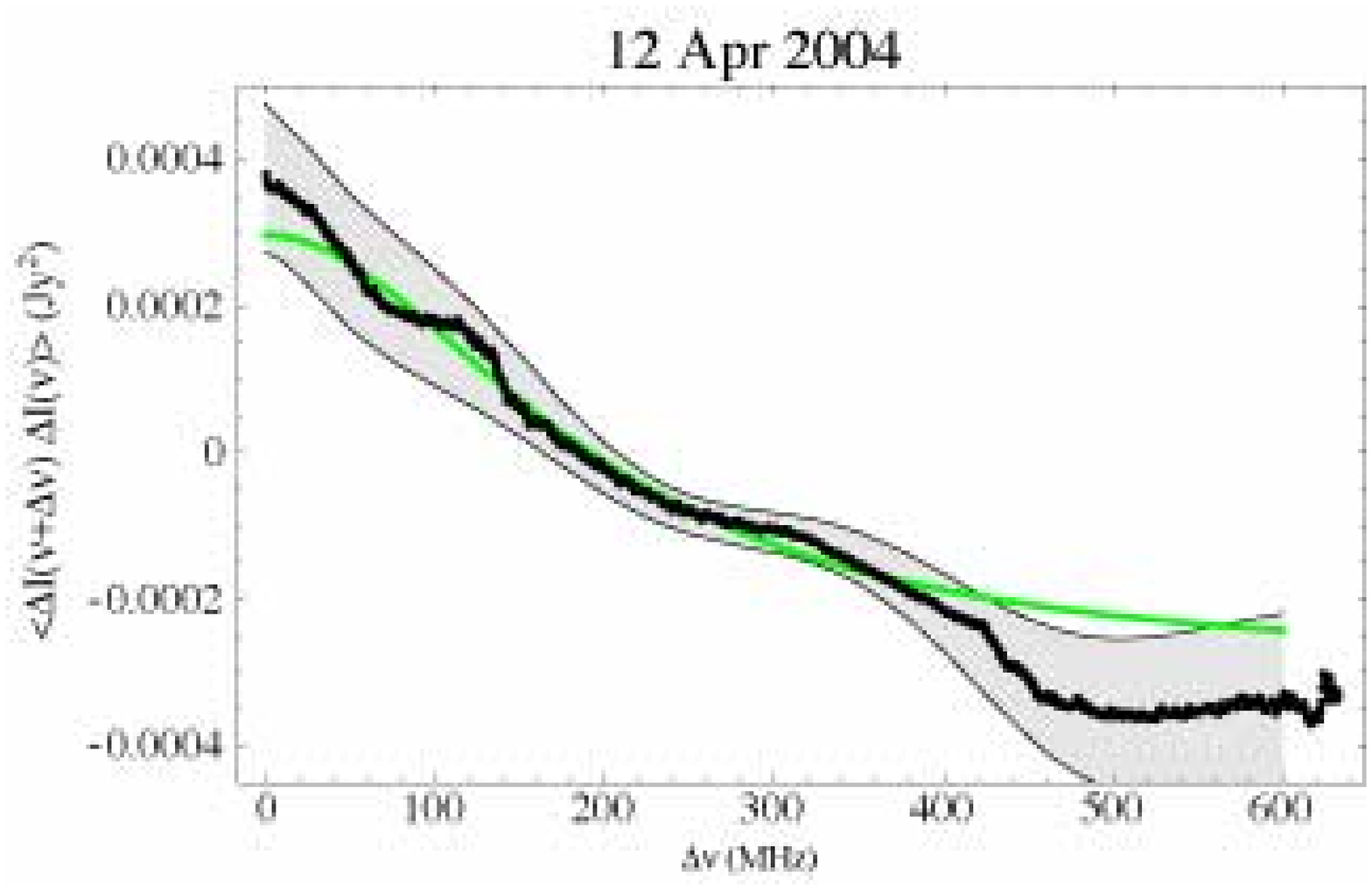,width=110mm}} 
\centerline{\psfig{file=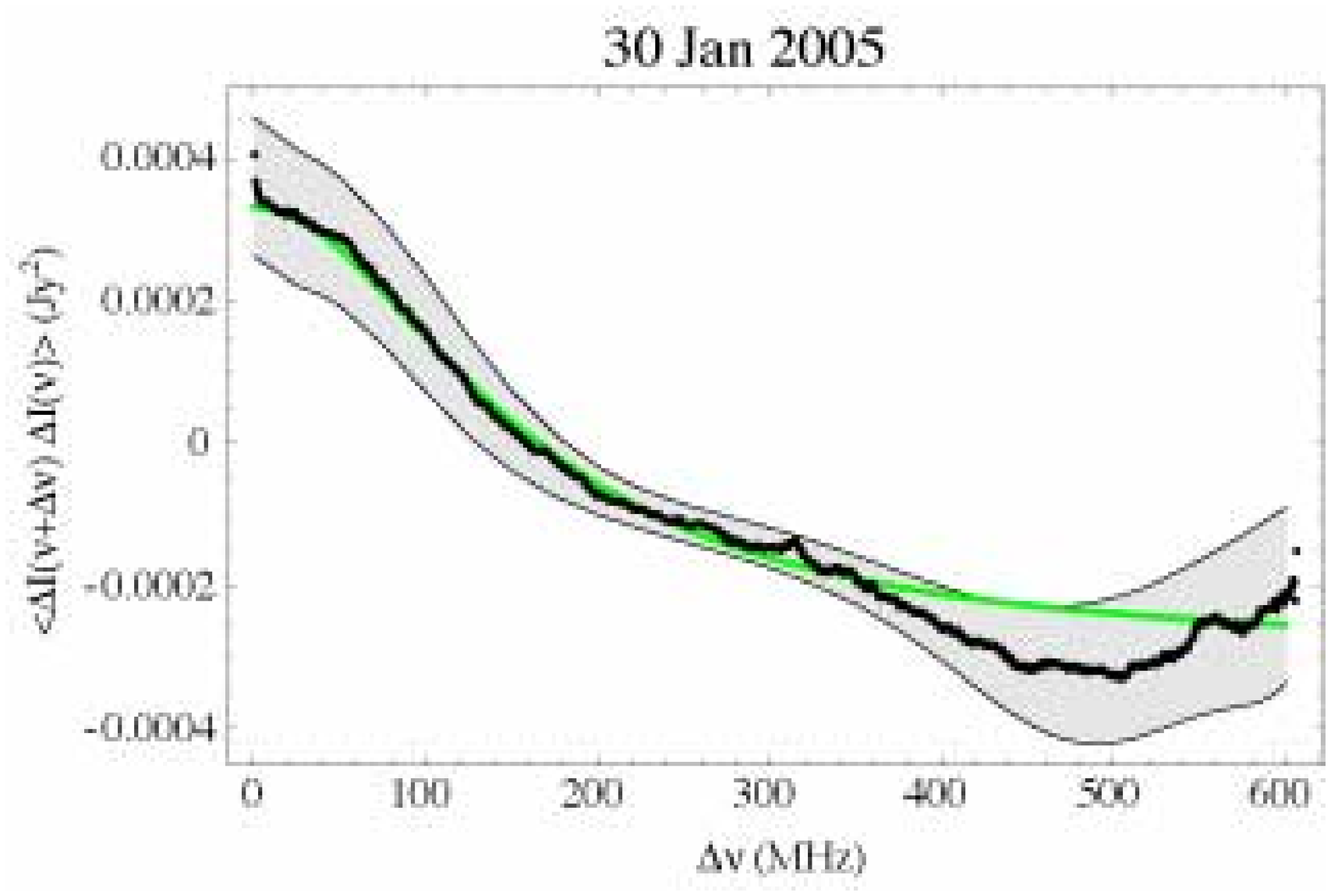,width=110mm}} 
\caption{Fits to the spectral decorrelation characteristics of the diffractive scintillation.  The grey error regions incorporate both uncertainties due to the finite number of diffractive scintles incorporated in the fit and uncertainties in the mean spectrum of the diffractively scintillating component.   The green and blue lines represent, respectively, the best fitting spectral decorrelation functions expected from a thin phase changing screen and from an extended scattering medium. Where only one line is visibile the fits are identical.  The plots shown here are the averages of the autocorrelations from $\approx 720$ time slots for each observation.  The observed decorrelation bandwidths (half-width at half maximum) are 170, 192 and 156\,MHz for the three epochs respectively.
} \label{ACFFigs}
\end{figure*}

\appendix

\section{Determination of errors in the spectral autocorrelation function} \label{AppErrs}
The finite number of scintles measured during each observation must be taken into account when comparing the spectral autocovariance curves shown in Fig.\,\ref{ACFFigs} to the theoretical models, which represent an average over the ensemble of all possible scintles.

The deviation in the sample autocovariance function from the ensemble-average autocovariance can be characterized in terms of the the variance in the sample autocovariance function, $C (\tau)$, given by (\cite{Jenkins}, eq. 5.3.21)
\begin{eqnarray}
{\rm var}[C (\tau)] = \frac{1}{(T- | \tau |)^2} \int_{-(T-\tau)}^{T-\tau} (T-\tau-r) [\gamma(r)^2 - \gamma(r+\tau) \gamma(r-\tau) ] \, dr,  \label{Cerrs} \nonumber \\
\end{eqnarray}
where $T$ is the length of the sample dataset and $\gamma(\tau)$ is the ensemble-average autocovariance.  

We wish to determine the error in the spectral autocovariance of a single time slot.  The ensemble average spectral autocovariance is unknown but, in order to determine the error in $C_\nu$, we approximate its form using the best-fitting thin screen autocovariance function.   The error for a single time channel is determined by numerically integrating this model in equation (\ref{Cerrs}), with $T$ set to the  total bandwidth spanned during the observation.   

The final spectral autocorrelations shown in Figure \ref{ACFFigs} represent an averages over many diffractive scintles observed during a 12-hour duration.  We assume that the autocovariance estimates of the individual diffractive scintles are statistically independent, so that the errors deduced from eq. (\ref{Cerrs}) are reduced by the square root of the number of diffractive scintles contributing to $C_\nu$. This is estimated by comparing the observing duration to the diffractive timescale.

These sampling errors are combined linearly with spectral misestimation errors (see \S\ref{Spectral}) to derive the error regions depicted in Fig.\,\ref{ACFFigs}.  Sampling errors dominate at all spectral lags.


\begin{thebibliography}{}


\bibitem[Baars et al. 1977]{Baars} Baars, J.W.M., Genzel, R., Pauliny-Toth, I.I.K. \& Witzel, A. 1977, A\&A, 61, 99

\bibitem[Begelman, Ergun \& Rees 2005]{Begelman0} Begelman, M.C., Ergun, R.E. \& Rees, M.J. 2005, \apj, 625, 51

\bibitem[Begelman, Rees \& Sikora 1994]{Begelman} Begelman, M.C., Rees, M.J., Sikora, M. 1994, Ap.J., 429, L57


\bibitem[Bignall et al. 2003]{Bignall} Bignall, H.E., Jauncey, D.L., Lovell, J.E.J., Tzioumis, A.K., Kedziora-Chudczer, L., Macquart, J.-P., Tingay, S.J., Rayner, D.P. \& Clay, R.W. 2003, ApJ, 585, 653

\bibitem[Chashei \& Shishov 1976]{Chashei} Chashei, I.V. \& Shishov, V.I. 1976, Soviet Astronomy, 20, 13

\bibitem[Coles et al. 1987]{Coles} Coles, W.A., Frehlich, R.G., Rickett, B.J. \& Codona, J.L. 1987, Ap.J., 315, 666

\bibitem[Codona \& Frehlich 1987]{Codona} Codona, J.L. \& Frehlich R.G. 1987, {Radio Science}, {22}, 469

\bibitem[Condon \& Backer 1975]{Condon} Condon, J.J. \& Backer, D.C. 1975, Ap.J., 197, 31

\bibitem[Coppi, Blandford \& Rees 1993]{Coppi} Coppi, P., Blandford, R.D. \& Rees, M.J. 1993, MNRAS, 262, 603


\bibitem[Dennett-Thorpe \& de Bruyn 2000]{DennettThorpe00} Dennett-Thorpe, J. \& de Bruyn, A.G. 2000,  Ap.J., 529, L65

\bibitem[Dennett-Thorpe \& de Bruyn 2002]{DennettThorpe02} Dennett-Thorpe, J. \& de Bruyn, A.G. 2002, Nature, 415, 57

\bibitem[Dennett-Thorpe \& de Bruyn 2003]{DennettThorpe03} Dennett-Thorpe, J. \& de Bruyn, A.G 2003, A\&A, 404, 113

\bibitem[Dennison \& Condon 1981]{Dennison} Dennison, B. \& Condon, J.J. 1981, ApJ, 246, 91

\bibitem[Ewing et al. 1970]{Ew70} Ewing, M.S., Batchelor, R.A., Friefeld, R.D., Price, R.M. \& Staelin, D.H. 1970, ApJ, 162, L169

\bibitem[Gedalin \& Eichler 1993]{Gedalin} Gedalin, M. \& Eichler, D. 1993, ApJ, 406, 629

\bibitem[Goodman \& Narayan 1985]{Goodman} Goodman, J. \& Narayan, R. 1985, 
MNRAS, 214, 519

\bibitem[Gwinn et al. 1998]{Gwinn} Gwinn, C.R., Britton, M.C., Reynolds, J.E., Jauncey, D.L., King, E.A., McCulloch, P.M., Lovell, J.E.J., Preston, A. 1998, ApJ, 505, 928

\bibitem[Heeschen 1984]{Heeschen} Heeschen, D.S. 1984, A.J., {89}, 1111 

\bibitem[Jauncey et al. 2000]{Jauncey00} Jauncey, D.L., Kedziora-Chudczer, L.L., Lovell, J.E.J., et al. 2000, in Astrophysical Phenomena Revealed by Space VLBI, Insitute of Space and Astronautical Science, p.147

\bibitem[Jauncey \& Macquart 2001]{JaunceyM} Jauncey, D.L. \& Macquart, J.-P. 2001, A\&A, 370, L9

\bibitem[Jenkins \& Watts 1968]{Jenkins} Jenkins, G.M. \& Watts, D.G., ``Spectral Analysis and its Applications'', Holden-Day 1968 

\bibitem[Kedziora-Chudczer et al. 1997]{Kedziora-Chudczer} Kedziora-Chudczer, L., Jauncey, D.L., Wieringa, M.H., Walker, M.A., Nicolson, G.D., Reynolds, J.E. \& Tzioumis, A.K. 1997, Ap.J., 490, L9

\bibitem[Kellerman \& Pauliny-Toth 1969]{Kellerman} Kellerman, K.I. \& Pauliny-Toth, I.I.K. 1969, {Ap.J.}, {155}, L71


\bibitem[Levinson \& Blandford 1995]{Levinson} Levinson, A. \& Blandford, R. 1995, \mnras, 274, 717

\bibitem[Lovell et al. 2003]{Lovell} Lovell, J.E.J., Jauncey, D.L., Bignall, H.E., Kedziora-Chudczer, L., Macquart, J.-P., Rickett, B.J. \& Tziomis 2003, AJ, 126, 1699

\bibitem[Luo \& Melrose 1995]{Luo} Luo, Q. \& Melrose, D.B. 1995, ApJ, 452, 346

\bibitem[Macquart, de Bruyn \& Dennett-Thorpe 2003]{MacquartParis} Macquart, J.-P., de Bruyn, A.G. \& Dennett-Thorpe, J., 2002, in ``Active Galactic Nuclei: from Central Engine to Host Galaxy'', ASP Conf. Series vol 290, p349.

\bibitem[Macquart et al. 2000]{Macquart00} Macquart, J.-P., Kedziora-Chudczer, L., Rayner, D.P. \& Jauncey, D.L. 2000, Ap.J.,  538, 623

\bibitem[Melrose 1991]{Melrose91} Melrose, D.B. 1991, ARA\&A, 29, 31

\bibitem[Mutel \& Lestrade 1990]{Mutel} Mutel, R.L. \& Lestrade, J.-F. 1990, Ap.J.,  349, L47

\bibitem[Narayan 1992]{Narayan} Narayan, R. 1992, Phil. Trans. R. Soc. Lond. A, 341, 151

\bibitem[Papoulis 1991]{Papoulis} Papoulis, A., Probability, random variables, and stochastic processes,  3rd ed., New York: McGraw Hill, 1991

\bibitem[Quirrenbach et al. 2000]{Quirrenbach00} Quirrenbach, A., Kraus, A., Witzel, A., Zensus, J.A., Peng, B., Risse, M., Krichbaum, T.P., Wegner, R. \& Naundorf, C.E. 2000, A\&A Supp., 141, 221

\bibitem[Quirrenbach et al. 1991]{Bonngp2} Quirrenbach, A., Witzel, A., Wagner, S., Sanchez-Pons, F., Krichbaum, T.P., Wegner, R., Anton, K., Erkens, U., Haehnelt, M., Zensus, J.A. \& Johnston, K.J. 1991, Ap.J., 372, L71

\bibitem[Readhead 1994]{Readhead} Readhead, A.C.S. 1994, {Ap.J.}, {426}, 51 


\bibitem[Rickett 1970]{Rickett70} Rickett, B.J. 1970, M.N.R.A.S., 150, 67

\bibitem[Rickett et al. 2001]{Rickett01} Rickett, B.J., Witzel, A., Kraus, A., Krichbaum, T.P. \& Qian, S.J., 2001, \apj, 550, L11

\bibitem[Rickett, Kedziora-Chudczer \& Jauncey 2002]{Rickett02} Rickett, B.J., Kedziora-Chudczer, L. \& Jauncey, D.L. 2002, Ap.J., 581, 103

\bibitem[Spangler \& Cordes 1998]{Spangler} Spangler, S.R. \& Cordes, J.M. 1998, Ap.J., 505, 766


\bibitem[Witzel et al. 1986]{Bonngp1} Witzel, A., Heeschen, D.S., Schalinski, C., Krichbaum, Th. 
1986, Mit.A.G., 65, 239

\end{thebibliography}
\end{document}